\documentclass{aa}

\usepackage{graphicx}
\usepackage{txfonts}
\usepackage{natbib}     
\usepackage{xcolor} 
\usepackage[para]{threeparttable} 
\usepackage{float}
\usepackage{threeparttable} 
\usepackage{placeins} 
\usepackage{multirow} 
\usepackage{tablefootnote}

\usepackage{caption}
\usepackage{subcaption}

\usepackage{natbib,twoopt} \usepackage[breaklinks=true]{hyperref}

\bibpunct{(}{)}{;}{a}{}{,} 

\DeclareRobustCommand{\FiftyOnePegb}{51\,Peg\,b\xspace}

\begin{document}

\title{\FiftyOnePegb revisited with VLT/CRIRES$^+$}
\subtitle{Constraints on atmospheric thermal structure, chemical composition, and an alternative orbital solution}

\author{
    D.~Cont \inst{\ref{instLMU}, \ref{instExzO}}
    \and L.~Nortmann \inst{\ref{instIAG}}
    \and F.~Lesjak \inst{\ref{instIAG}, \ref{instAIP}}
    \and U.~Seemann \inst{\ref{instESO}}
    \and L.~D.~Nielsen \inst{\ref{instLMU}}
    \and B.~Wechselberger \inst{\ref{instLMU}}
    \and F.~Yan \inst{\ref{instUSTC}}
    \and S.~Liu \inst{\ref{instUSTC}} 
    \and A.~Reiners \inst{\ref{instIAG}}    
    \and A.~Lavail \inst{\ref{instIRAP}}
    \and D.~Shulyak \inst{\ref{instIAA}}    
    \and A.~Hatzes \inst{\ref{instTLS}}
    \and N.~Piskunov \inst{\ref{instUppsala}}
    \and L.~Boldt-Christmas \inst{\ref{instUppsala}}
    \and S.~Czesla \inst{\ref{instTLS}}
    \and E.~W.~Guenther \inst{\ref{instTLS}} 
    \and O.~Kochukhov \inst{\ref{instUppsala}}
    \and T.~Marquart \inst{\ref{instUppsala}}
    \and E.~Nagel \inst{\ref{instIAG}}
    \and A.~D.~Rains \inst{\ref{instUppsala}, \ref{instIAC}}
    \and M.~Rengel \inst{\ref{instMPS}}
}

\institute{
    Universit\"ats-Sternwarte, Ludwig-Maximilians-Universit\"at M\"unchen, Scheinerstrasse 1, 81679 M\"unchen, Germany\label{instLMU}\\ 
    \email{david.cont@lmu.de}
    \and
    Exzellenzcluster Origins, Boltzmannstrasse 2, 85748 Garching bei M\"unchen, Germany\label{instExzO}
    \and
    Institut f\"ur Astrophysik und Geophysik, Georg-August-Universit\"at G\"ottingen, Friedrich-Hund-Platz 1, 37077 G\"ottingen, Germany\label{instIAG} 
    \and
    Leibniz-Institut für Astrophysik Potsdam (AIP), An der Sternwarte 16, 14482 Potsdam, Germany\label{instAIP}
    \and
    European Southern Observatory, Karl-Schwarzschild-Str. 2, 85748 Garching bei M\"unchen, Germany\label{instESO}
    \and
    Department of Astronomy, University of Science and Technology of China, Hefei 230026, People’s Republic of China\label{instUSTC}
    \and
    Institut de Recherche en Astrophysique et Plan\'etologie, Universit\'e de Toulouse, CNRS, IRAP/UMR 5277,
    14 avenue Edouard Belin, F-31400, Toulouse, France\label{instIRAP} 
    \and 
    Instituto de Astrof{\'i}sica de Andaluc{\'i}a (IAA-CSIC), Glorieta de la Astronom{\'i}a s/n, 18008 Granada, Spain\label{instIAA}
    \and
    Th\"uringer Landessternwarte Tautenburg, Sternwarte 5, 07778 Tautenburg, Germany\label{instTLS}
    \and
    Department of Physics and Astronomy, Uppsala University, Box 516, 75120 Uppsala, Sweden\label{instUppsala} 
    \and
    Instituto de Astrof{\'i}sica, Pontificia Universidad Cat{\'o}lica de Chile, Av. Vicu{\~n}a Mackenna 4860, 782-0436 Macul, Santiago, Chile\label{instIAC}    
    \and
    Max-Planck-Institut für Sonnensystemforschung, Justus-von-Liebig-Weg 3, 37077 G\"ottingen, Germany\label{instMPS}
}

\date{Received 14 November 2025 / Accepted 18 May 2026}


\abstract
{So far, the majority of high-resolution spectroscopy studies on exoplanet atmospheres have focused on transiting planetary systems. Consequently, the atmospheres of non-transiting exoplanets remain poorly explored, and the potential of high-resolution spectroscopy for determining key planetary parameters beyond their atmospheric properties has not been fully exploited. 
We obtained high-resolution emission spectra of the non-transiting hot Jupiter \FiftyOnePegb in the K band with VLT/CRIRES$^+$ to study its atmospheric thermochemical and dynamical structure and derive additional planetary parameters from the Doppler shift of its spectral lines. 
Using the cross-correlation technique, we confirmed the spectral signature of H$_2$O in the planetary emission spectrum and a non-inverted atmospheric temperature profile. An indication was also found for the presence of atmospheric CO, although with a signal strength below the threshold for significant detection. The atmospheric chemical and thermal conditions were quantitatively constrained by use of a Bayesian retrieval framework, which yielded a high metallicity value ($2.63_{-0.93}^{+1.00}$\,dex) suggestive of chemical quenching, a solar C/O ratio ($0.54_{-0.23}^{+0.18}$), and a temperature profile in the upper atmosphere in line with the expected planetary equilibrium temperature. Moreover, we measured an excess of spectral line broadening, which suggests the presence of atmospheric dynamics in the atmosphere of \FiftyOnePegb. The planetary signal was recovered at an orbital semi-amplitude of $102.8_{-9.1}^{+8.3}$\,km\,s$^{-1}$, differing from previous high-resolution measurements. From the obtained orbital semi-amplitude, we further derived the surface gravity ($3.15 \pm 0.12$\,$\log\mathrm{cgs}$), mass ($0.61_{-0.05}^{+0.06}$\,$M_\mathrm{Jup}$), and orbital inclination ($49.8_{-5.7}^{+5.8}$\,deg) of \FiftyOnePegb. The mass and inclination values differ from those reported in earlier works. Overall, this study provides new constraints on the thermochemical and dynamical properties of \FiftyOnePegb's atmosphere and demonstrates the potential of high-resolution emission spectroscopy for measuring parameters related to the planetary orbital motion.}


\maketitle

%

\section{Introduction}

Hot Jupiters are gas giant exoplanets on close-in orbits around their host stars, with orbital periods on the order of a few days. The intense stellar irradiation they receive leads to elevated equilibrium temperatures ($T_\mathrm{eq}$) of about 1500\,K, resulting in extended atmospheric scale heights and strong thermal emission fluxes. This makes these exoplanets particularly favorable targets for atmospheric characterization via transmission and emission spectroscopy. Hot Jupiters typically have circular orbits and are expected to exhibit synchronous rotation, both of which are caused by tidal circularization during the early evolution of these planetary systems \citep{Hut1981}.

Despite the intense stellar irradiation they receive, the atmospheres of hot Jupiters are sufficiently cool to allow the presence of significant abundances of molecular species, as most molecules thermally dissociate only at $T_\mathrm{eq}$ beyond approximately 2000\,K \citep{Parmentier2018}. Both space- and ground-based observations have identified CO and H$_2$O as dominant contributors to the spectra of hot Jupiters at near-infrared wavelengths \citep[e.g.,][]{Snellen2010, Wakeford2018}. These species contain a substantial fraction of the atmospheric carbon and oxygen and therefore serve as key tracers of a planet's carbon-to-oxygen (C/O) ratio, which is commonly used to probe exoplanet formation pathways \citep[e.g.,][]{Oeberg2011, Mordasini2016, Williams2025}. Beyond CO and H$_2$O, numerous other molecular species have been discovered in the atmospheres of hot Jupiters, such as CO$_2$, CH$_4$, and SO$_2$ \citep[e.g.,][]{Giacobbe2021, Carter2024, Xue2024}. 

The opacities of molecular species are generally lower at visible than at infrared wavelengths. As a result, stellar radiation can penetrate efficiently into hot Jupiter atmospheres, while the thermal reemission of the absorbed energy at infrared wavelengths is comparatively less efficient. This effect causes an increase in temperature with atmospheric pressure in hot Jupiter atmospheres, yielding a non-inverted temperature-pressure \mbox{($T$-$p$)} profile \citep[e.g.,][]{Fortney2008, Guillot2010}.

The thermal conditions in hot Jupiter atmospheres allow refractory elements to condense out of the gas phase, leading to the formation of clouds. The pressure level at which these clouds are located depends on a planet's $T$-$p$ profile and atmospheric composition \citep{Helling2019}. In addition, photochemical reactions in the upper atmospheres of hot Jupiters result in complex molecular species that form solid particles referred to as hazes \citep{Gao2021}. Clouds and hazes can alter the spectra of hot Jupiter atmospheres. For instance, these aerosols can obscure the spectral features of the underlying atmospheric layers. Moreover, clouds and hazes affect the atmospheric $T$-$p$ profile through their strong influence on a planet's albedo as well as by inducing cooling and warming effects above and below the cloud deck, respectively \citep{Heng2012, Gao2021}.

General circulation models predict the presence of strong dynamical processes in the atmospheres of hot Jupiters. These large-scale atmospheric gas flows are classified into two regimes: dayside to nightside winds and super-rotating equatorial jet streams, both of which can reach velocities of several km\,s$^{-1}$ \citep[e.g.,][]{MillerRicciKempton2012, Showman2013, TanKomacek2019}. Detections of Doppler offsets relative to the planetary rest frame and broadening in the spectral line profiles have confirmed the presence of such global-scale wind patterns in observational studies. For instance, \cite{Nortmann2025} revealed the presence of a fast super-rotating jet stream in the atmosphere of the hot Jupiter WASP-127b, while \cite{Seidel2025} identified dayside to nightside winds and jet stream circulation at different pressure levels in the atmosphere of WASP-121b.

High-resolution spectroscopy has proven to be a powerful tool for probing the chemical composition, thermal structure, and atmospheric dynamics of hot Jupiters \citep[e.g.,][]{Snellen2010, Birkby2013, Line2021, Rafi2024, Kanumalla2024, Finnerty2025}. Beyond providing detailed insights into atmospheric conditions, this technique enables the derivation of key orbital and planetary parameters, such as the planetary mass and orbital inclination. For transiting hot Jupiters, determination of the planetary mass is relatively straightforward by using the radial velocity (RV) method, which traces the periodic Doppler shift of the spectral lines of an exoplanet's host star. In this specific case, the orbital inclination, which is degenerate with planetary mass in the RV signal, is known from the transit geometry. However, RV measurements of the host star are unable to break the degeneracy between planetary mass and orbital inclination for non-transiting hot Jupiters. This limitation can be overcome by measuring the thermal emission signal from hot Jupiter atmospheres, which allows the planetary orbit to be traced directly. So far, this approach has been successfully applied to the study of the planetary and orbital parameters of a few bright, non-transiting hot Jupiters, including $\tau$\,Boo\,b \citep{Brogi2012}, HD\,179949\,b \citep{Brogi2014}, and \FiftyOnePegb \citep[e.g.,][]{Brogi2013, Birkby2017}.

\FiftyOnePegb was the first exoplanet ever detected orbiting a main-sequence star \citep{MayorQueloz1995} and the first hot Jupiter ever discovered. As the planet does not transit in front of the host star from the perspective of the Earth-bound observer, its atmospheric properties cannot be probed via transmission spectroscopy. Atmospheric characterization is therefore only possible via the analysis of reflected stellar light or the planet's thermal emission spectrum. A number of studies have attempted to detect reflected stellar light from \FiftyOnePegb. \cite{Martins2015} analyzed HARPS-N observations of the planet in the visible wavelength range and reported tentative evidence of a reflection signal. However, this signal could only be explained by assuming an elevated geometric albedo and an inflated planetary radius. Using the same dataset, \cite{BorraDeschatelets2018} also found evidence of reflected stellar light in the spectrum of \FiftyOnePegb. In contrast, a reanalysis of these observations by \cite{DiMarcantonio2019} yielded a non-detection of the reflected stellar signal. A subsequent study by \cite{Scandariato2021}, using the same data along with additional HARPS-N measurements, derived a geometric albedo below 0.15 and concluded that no reflected stellar light could be detected. More recently, \cite{Spring2022} reported the absence of significant reflected light in HARPS-N data. While robust detections of reflected stellar light remain elusive, the thermal emission spectrum of \FiftyOnePegb has been successfully detected at high spectral resolution in the infrared, providing a direct window into its atmospheric composition and thermal structure. \cite{Brogi2013} first identified CO and H$_2$O spectral lines in the planet’s emission spectrum using K band observations with the old CRIRES. Later, \cite{Birkby2017} reported detection of H$_2$O in the L band using the same instrument. Additionally, \cite{Chiavassa2019} confirmed the presence of CO and H$_2$O features by reanalyzing the data from \cite{Brogi2013} with improved methods for removing stellar contamination. 

This present study provides the first spectroscopic characterization of the benchmark hot Jupiter \FiftyOnePegb using Bayesian inference methods and data from the upgraded CRIRES$^+$ instrument, offering a robust quantitative analysis of its atmospheric thermal structure, chemical composition, and dynamical processes. This target was identified as one of the most suitable systems for high-resolution spectroscopic studies of exoplanet atmospheres by \cite{Shulyak2019}. All parameters of the 51\,Peg system used in this work are summarized in the Table~\ref{table:planet-parameters}. The paper is organized as follows: In Sects.~\ref{sec:observations_and_data_reduction} and \ref{sec:pre_processing_the_spectra}, we describe the observations, data reduction, and preprocessing procedures. In Sect.~\ref{sec:updated_ephemeris_from_RV_analysis}, we update the ephemeris of the 51\,Peg system. The methodology used to identify spectral features in the planet’s emission spectrum and the Bayesian framework employed to constrain its atmospheric properties are detailed in Sect.~\ref{sec:atmospheric_analysis_methods}. The results are presented and discussed in Sect.~\ref{sec:results_and_discussion}, and our main conclusions are summarized in Sect.~\ref{sec:conclusions}.

\begin{table}
        \caption{Parameters of the \object{51\,Peg} system.}
        \label{table:planet-parameters}
        \centering 
        \renewcommand{\arraystretch}{1.4}
        \begin{threeparttable}
                \begin{tabular}{l@{\hspace{-0mm}}c}
                \hline\hline
                \noalign{\smallskip}
                Parameter & Value \\
                \noalign{\smallskip}
                \hline 
                \noalign{\smallskip}
                \textit{Planet} &  \\ 
                \noalign{\smallskip}
                Time of conjunction ($T_\mathrm{C}$)      & $2456326.9323 \pm 0.0013$\,BJD$_\mathrm{TDB}$ \\ 
                Orbital period ($P_\mathrm{orb}$)         & $4.2307966 \pm 0.0000027$\,d \\
                Mass ($M_\mathrm{p}$)                     & $0.61_{-0.05}^{+0.06}$\,$M_\mathrm{Jup}$ \\          
                RV semi-amplitude ($K_\mathrm{p}$)        & $102.8_{-9.1}^{+8.3}$\,km\,s$^{-1}$ \\    
                Orbital inclination ($i$)                 & $49.8_{-5.7}^{+5.8}$\,deg \\                       
                Surface gravity ($\log{g}$)               & $3.15 \pm 0.12$\,$\log{\mathrm{cgs}}$ \\
                Semi-major axis ($a$)                     & 0.052\,au\,$^{\mathrm{(a)}}$ \\
                
                \noalign{\smallskip} \hline \noalign{\smallskip}
                \textit{Star} &  \\  
                \noalign{\smallskip}
                Radius ($R_*$)                            & $1.19 \pm 0.03$\,$R_\mathrm{\sun}$\,$^{\mathrm{(a)}}$ \\ 
                Mass ($M_*$)                              & $1.07 \pm 0.05$\,$M_\mathrm{\sun}$\,$^{\mathrm{(a)}}$ \\ 
                RV semi-amplitude ($K_*$)                 & $55.77 \pm 0.15$\,m\,s$^{-1}$ \\ 
                Systemic velocity ($\varv_\mathrm{sys}$)  & $-33.1392 \pm 0.00010$\,km\,s$^{-1}$ \\ 
                Effective temperature ($T_\mathrm{eff}$)  & $5761 \pm 95$\,K\,$^{\mathrm{(a)}}$ \\
                
                \noalign{\smallskip}
                \hline                    
                \end{tabular}
                \vspace{0.3cm}
                \begin{tablenotes}
                \textbf{Notes.} $^{\mathrm{(a)}}$ Values taken from \cite{Rosenthal2021}; all other values derived in this work. For $\varv_\mathrm{sys}$, we report the value derived from the RV analysis of the ESPRESSO data, given the instrument’s absolute velocity calibration.
                \end{tablenotes} 
        \end{threeparttable}
\end{table}

%

\section{Observations and data reduction}
\label{sec:observations_and_data_reduction}

We observed the thermal emission spectrum of \FiftyOnePegb on 6 October 2024 between 01:17\,UT and 04:10\,UT with the CRIRES$^+$ instrument at the Unit Telescope 3 of the Very Large Telescope \citep{Dorn2023}. The observations cover orbital phases 0.52--0.55, which are close to superior conjunction, when the planetary dayside is most favorably aligned with the observer's line of sight. A total of 130 consecutive spectra were collected. The adaptive optics (AO) system was disabled during the acquisition of the first 28 spectra due to passing clouds; the remaining 102 spectra were obtained under improved meteorological conditions that allowed for AO usage. The AO system was disabled under cloud coverage because its wavefront sensor has a limited dynamic range that cannot be adjusted during observations. Rapid flux variations from passing clouds can cause the flux to exceed this range, which could damage the wavefront sensor. We verified that including data without AO does not affect our results. Both the cross-correlation and retrieval analyses in Sect.~\ref{sec:results_and_discussion} yield consistent outcomes with or without these spectra. Each spectrum was taken with DIT\,=\,15\,s and NDIT\,=\,4, resulting in a total integration time of 60\,s per frame. Splitting each frame into four sub-integrations was required to prevent saturation of the detector pixels, given the bright host star of our target. 

We used an \mbox{AABBBBAA} nodding pattern, which consists of observing the target at two different slit positions (A and B), to facilitate the removal of the sky background during the subsequent data reduction steps. The slit width was set to 0.2$^{\prime\prime}$, which corresponds to a nominal resolving power of $R$\,$\sim$\,100\,000 under the condition of a homogeneous slit illumination. However, during our observations, the seeing conditions were very good, such that the AO system -- when active -- could achieve a point spread function (PSF) smaller than the slit. The reduced PSF width increased the spectral resolution from $R$\,$\sim$\,100\,000 (28 spectra with no AO) to $R$\,$\sim$\,117\,000 (102 spectra with AO). 
We determined the spectral resolution using {\tt molecfit} \citep{Smette2015} via the model kernel full width at half maximum (FWHM) stored in the BEST\_FIT\_PARAMETERS output file. The FWHM in detector pixels was first converted to wavelength and then to spectral resolution. For the AO assisted spectra affected by super-resolution, the FWHM corresponds to approximately 2.6 pixels. This exceeds the two-pixel minimum required for Nyquist sampling, ensuring adequate sampling of the instrumental profile. 
The data were acquired in the K2166 wavelength setting, which offers a wavelength coverage between 1921\,nm and 2472\,nm. This setting consists of seven echelle orders, each divided by two narrow gaps between detectors, yielding a total of 21 wavelength segments. The airmass decreased from 1.5 to a minimum of 1.4 before increasing to 1.6; the mean signal-to-noise (S/N) ratio per detector pixel was 330 and 485 for spectra obtained without and with AO, respectively.

We used the data reduction pipeline {\tt cr2res} to extract the one-dimensional spectra from the raw frames\footnote{\url{https://www.eso.org/sci/software/pipe_aem_table.html}}. This procedure includes correction for dark frames and flat fields, removal of bad pixels, and calculation of a wavelength solution. The integration time of the individual frames was shorter than the timescale over which \FiftyOnePegb's RV changes significantly with respect to the observer (60\,s compared to $\sim400$\,s to smear the planetary signal over 1\,km\,s$^{-1}$, which is on the order of one CRIRES$^+$ detector pixel). Therefore, we combined adjacent AA and BB frame pairs into one A frame and one B frame, respectively. This limited the subsequent data reduction to the analysis of a 65 frame ABBA nodding sequence, with an integration time of 120\,s per frame. 

The wavelength solution for the A and B spectra, as calculated by the {\tt cr2res} pipeline, is based on data collected during daytime calibrations. However, this wavelength solution can differ from that of the nighttime science data due to instrumental drifts or offsets between subsequent frames caused by changes in ambient or observing conditions (e.g., variations in temperature, pressure, or slit illumination). Therefore, we used {\tt molecfit} to refine the wavelength solution of our spectra. Following \cite{Cont2024}, we first derived wavelength solutions for the time averaged A and B spectra. Subsequently, we determined the offsets between the individual spectra and the corresponding A or B wavelength solution. To this end, we computed the cross-correlation between each individual spectrum and the respective A or B time averaged spectrum. Finally, we shifted each A and B spectrum to the rest frame of its time averaged spectrum. Figure~\ref{fig:wavelength-offset} shows how these procedures align the individual spectra to the refined wavelength solution. We treated the A~and~B spectra as two independent datasets throughout the data processing procedures in Sect.~\ref{sec:pre_processing_the_spectra}, and combined their information in Sect.~\ref{sec:atmospheric_analysis_methods}.

\section{Preprocessing the spectra}
\label{sec:pre_processing_the_spectra}

We normalized the extracted spectra of our time series to a common continuum level. To this end, we first computed the median of all spectra in the time series and fitted the resulting master spectrum with a second-order polynomial. Each individual spectrum was then divided by this master and fitted with a first-order polynomial. The final continuum correction was obtained by multiplying these two polynomial fits and dividing each spectrum by the result. The continuum-normalized spectra were then arranged chronologically in a two-dimensional array to form a spectral matrix. We removed outliers by 5$\sigma$ clipping and masked wavelength regions with deep telluric and stellar lines. In total, 19\% of the data points was masked. A more detailed description of these procedures can be found in \cite{Cont2024, Cont2025}.

Telluric and stellar contamination were removed with {\tt SYSREM} from the normalized spectral matrix \citep{Tamuz2005}. This iterative principle component analysis algorithm has facilitated the study of exoplanet atmospheres at high spectral resolution in recent years \citep[e.g.,][]{Birkby2013, Birkby2018}. We followed the procedure proposed by \cite{Gibson2022}, which involves dividing the spectral matrix by the median spectrum before applying the {\tt SYSREM} algorithm. This approach preserves the relative strengths of the planetary spectral lines and allows one to account for distortions introduced by the algorithm when applying a Bayesian retrieval framework in Sect.~\ref{sec:retrieval}. We ran {\tt SYSREM} for up to ten consecutive iterations, producing a residual spectral matrix for each iteration and wavelength segment. The residual spectral matrix is expected to contain only the planetary spectral signal, which is dominated by noise.

To determine the optimal number of {\tt SYSREM} iterations, we assessed the relative change in standard deviation of the residual spectral matrix from one iteration to the next. This approach has been successfully used in previous studies \citep[e.g.,][]{Parker2025, PelaezTorres2025}. The relative change in standard deviation is defined as  

\begin{equation}
     \Delta \sigma = \frac{^{(i-1)}\sigma - ^{(i)}\sigma}{^{(i-1)}\sigma}, 
     \label{eq:delta_sigma}
\end{equation}

where $^{(i-1)}\sigma$ and $^{(i)}\sigma$ denote the standard deviation of the residual spectral matrix before and after the $i$-th {\tt SYSREM} iteration across all wavelength segments. We adopted a conservative approach by determining a common optimal iteration number across all wavelength segments, rather than determining the optimal number for each segment individually. Segment-by-segment optimization has been shown to pose a risk of false-positive detections \citep[e.g.,][]{SanchezLopez2025}. The $\Delta \sigma$ value of our data plateaus after six iterations (Fig.~\ref{fig:sysrem}). This indicates that {\tt SYSREM} no longer identifies significant correlated trends beyond this number of iterations.

\section{Updated ephemeris from RV analysis}
\label{sec:updated_ephemeris_from_RV_analysis}

Accurate planetary ephemerides are essential for precise orbital phase assignment of spectroscopic data, which in turn allows for more reliable constraints on atmospheric properties. To update the ephemeris of \FiftyOnePegb, we conducted an RV analysis including publicly available data of 51\,Peg from Levay/APF \citep{Vogt2014} and VLT/ESPRESSO \citep{Pepe2021}, as well as the Lick/Hamilton, Keck/HIRES, OHP/ELODIE, and \mbox{ESO\,3.6-m\,Telescope/HARPS} RV datasets from \cite{Birkby2017}. We accessed the APF measurements made available in \cite{Rosenthal2021}, which consist of a total of 283 epochs with a typical RV uncertainty of 2.7\,m\,s$^{-1}$ spanning five years from 16 October 2013 to 14 October 2018. The ESPRESSO data was accessed through the ESO archive and reduced with the instrument data reduction pipeline version 3.3.1\footnote{\url{https://www.eso.org/sci/software/pipe_aem_table.html}}. We extracted the RV information via cross-correlating the spectra with a G2 binary mask using default pipeline settings. The 90 ESPRESSO RV data points were obtained on three different nights covering a baseline of 12 days in August 2019. The very high RV precision of 0.15\,m\,s$^{-1}$ allowed this data to add significant information content to the final fit. Together, the datasets we used span more than 25 years. For the final modeling of the data, we ensured all timestamps were in Barycentric Dynamical Time (BJD$_\mathrm{TDB}$).

\begin{figure}
        \centering
        \includegraphics[width=\columnwidth]{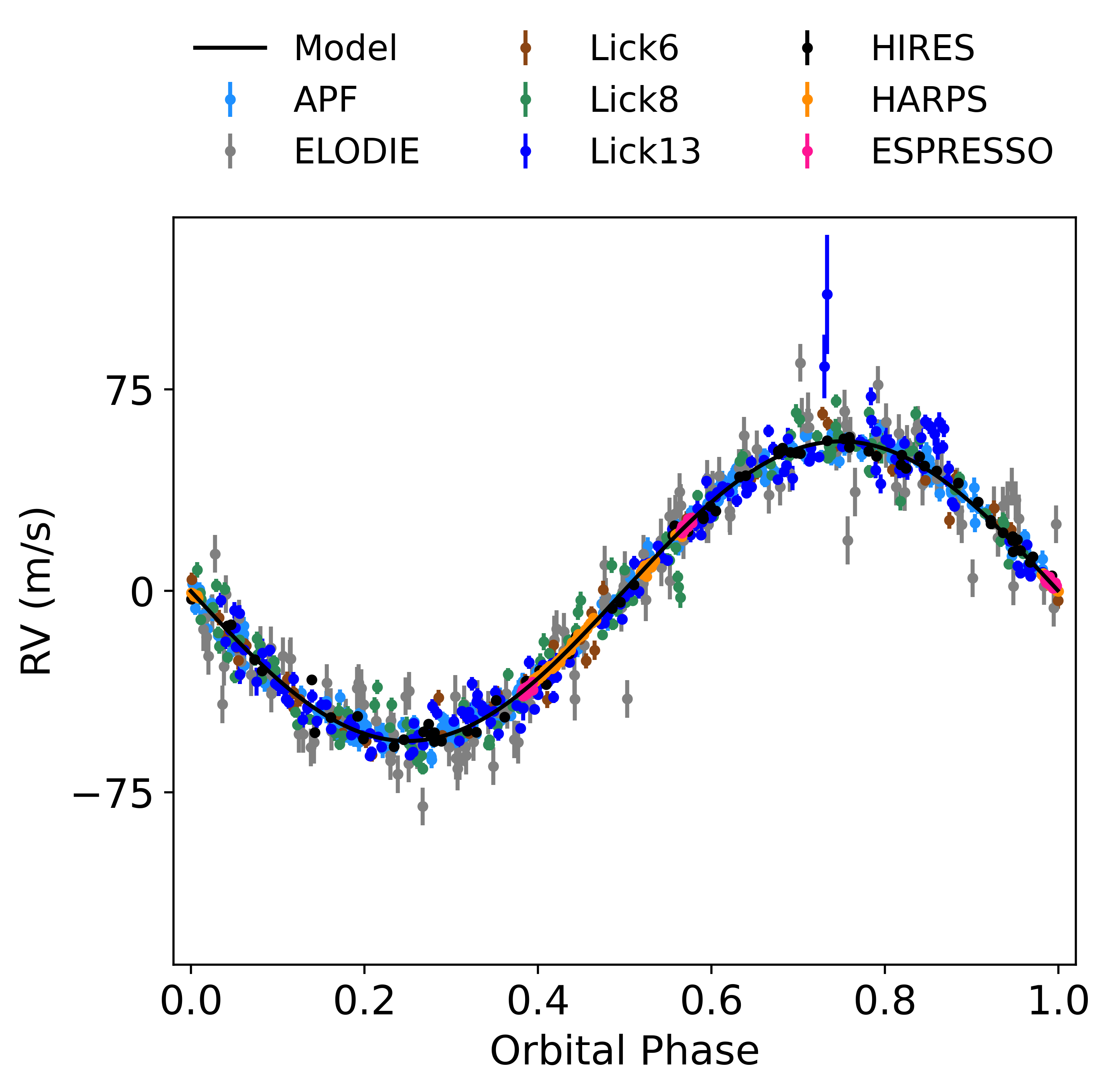}
        \caption{Radial velocity curve of 51\,Peg with the updated orbital solution. We show the phase-folded RV data points and the best-fit Keplerian model.}
        \label{fig:RV-analysis}
\end{figure}

To model the Keplerian orbital parameters, we used \texttt{PYANETI} \citep{Barragan2019} employing a sinusoid fit and Markov chain Monte Carlo (MCMC) sampler to assess the uncertainties. The datasets of the different instruments were treated independently to allow for offsets in systemic velocity and jitter terms. The Hamilton data were split up into three versions of the instrument to account for instrument interventions, as described in \cite{Birkby2017}. For the stellar parameters, we used the most recent literature results from \cite{Rosenthal2021}. 
In a first model run, we treated the orbital eccentricity as a free parameter. After obtaining a two-sigma upper eccentricity limit of $0.0063$, which is consistent with a circular orbit, we repeated the RV analysis with the parameter value fixed to zero. The results of this zero-eccentricity fit were then used for the subsequent analyses of \FiftyOnePegb’s atmospheric and planetary properties, based on CRIRES$^+$ observations (Sect.~\ref{sec:results_and_discussion}). Figure~\ref{fig:RV-analysis} shows all data points together with the best-fit Keplerian model, and Table~\ref{table:RV-parameters} lists the results of the RV analysis.

\begin{table}
        \caption{Parameters from the RV analysis of \object{51\,Peg}.}
        \label{table:RV-parameters}
        \centering 
        \renewcommand{\arraystretch}{1.4}
        \begin{threeparttable}
                \begin{tabular}{l@{\hspace{-5mm}}r}
                \hline\hline
                \noalign{\smallskip}
                Parameter & Value \\
                \noalign{\smallskip}
                \hline 
                \noalign{\smallskip}

                Time of conjunction ($T_C$)                     & $2456326.9323 \pm 0.0013 $\,BJD$_\mathrm{TDB}$ \\
                Orbital period ($P_\mathrm{orb}$)               & $4.2307966 \pm 0.0000027$\,d \\
                Eccentricity ($e$)                              & $< 0.0063$\,$^{\mathrm{(a)}}$ \\
                Stellar RV semi-amplitude ($K_*$)               & $55.77 \pm 0.15$\,m\,s$^{-1}$ \\
                Systemic velocity ($\varv_\mathrm{sys}$) APF    & $0.00042 \pm 0.00020 $\,km\,s$^{-1}$   \\
                $\varv_\mathrm{sys}$ ELODIE                     & $-33.2515 _{-0.00096 }^{+0.00097}$\,km\,s$^{-1}$ \\
                $\varv_\mathrm{sys}$ ESPRESSO                   & $-33.1392 \pm 0.00010 $\,km\,s$^{-1}$ \\
                $\varv_\mathrm{sys}$ Lick6                      & $0.01452_{-0.00087}^{+0.00088}$\,km\,s$^{-1}$   \\
                $\varv_\mathrm{sys}$ Lick8                      & $0.00525_{-0.00067}^{+0.00068}$\,km\,s$^{-1}$   \\
                $\varv_\mathrm{sys}$ Lick13                     & $0.02253 \pm 0.00047 $\,km\,s$^{-1}$   \\
                $\varv_\mathrm{sys}$ HIRES                      & $-0.00635 \pm 0.00034 $\,km\,s$^{-1}$  \\
                $\varv_\mathrm{sys}$ HARPS                      & $-33.15218 \pm 0.00012 $\,km\,s$^{-1}$ \\
                
                \noalign{\smallskip}
                \hline                    
                \end{tabular}
                \vspace{0.3cm}
                \begin{tablenotes}
                \textbf{Notes.} Systemic velocity ($\varv_\mathrm{sys}$) values around 0\,km\,s$^{-1}$ are measured relative to the rest frame of 51\,Peg, while $\varv_\mathrm{sys}$ values of approximately $-33$\,km\,s$^{-1}$ are measured relative to the barycentric rest frame. $^{\mathrm{(a)}}$ Value obtained with $e$ treated as a free parameter; all other values were derived with $e$ fixed to zero, as the free-eccentricity solution is consistent with a circular orbit.
                \end{tablenotes} 
        \end{threeparttable}
\end{table}

\section{Atmospheric analysis methods}
\label{sec:atmospheric_analysis_methods}

\subsection{Cross-correlation}
\label{sec:cross_correlation}

We used the radiative transfer code {\tt petitRADTRANS} to model emission spectra of \FiftyOnePegb's atmosphere \citep{Molliere2019}. In a preliminary cross-correlation analysis using the $T$-$p$ profile and chemical abundances from \cite{Brogi2013}, we achieved a significant detection of the planetary spectral lines in our data. We then applied the retrieval framework described in Sect.~\ref{sec:retrieval} to derive the atmospheric thermal and chemical properties (see Sect~\ref{sec:retrieval_of_the_atmospheric_properties}). The best-fit $T$-$p$ profile and volume mixing ratios (VMRs) of the considered chemical species were subsequently used to generate the model spectra corresponding to the cross-correlation analysis results presented in Sect.~\ref{sec:detection_of_the_planetary_spectral_lines}. The {\tt petitRADTRANS} code was provided with the spectral line lists of the chemical species under investigation. The CO line data were taken from \cite{Li2015}, and the H$_2$O data were taken from the POKAZATEL database \citep{Polyansky2018}. We generated spectra for CO and H$_2$O individually, as well as for the combination of the two species.

\begin{figure*}
        \centering
        \includegraphics[width=\textwidth]{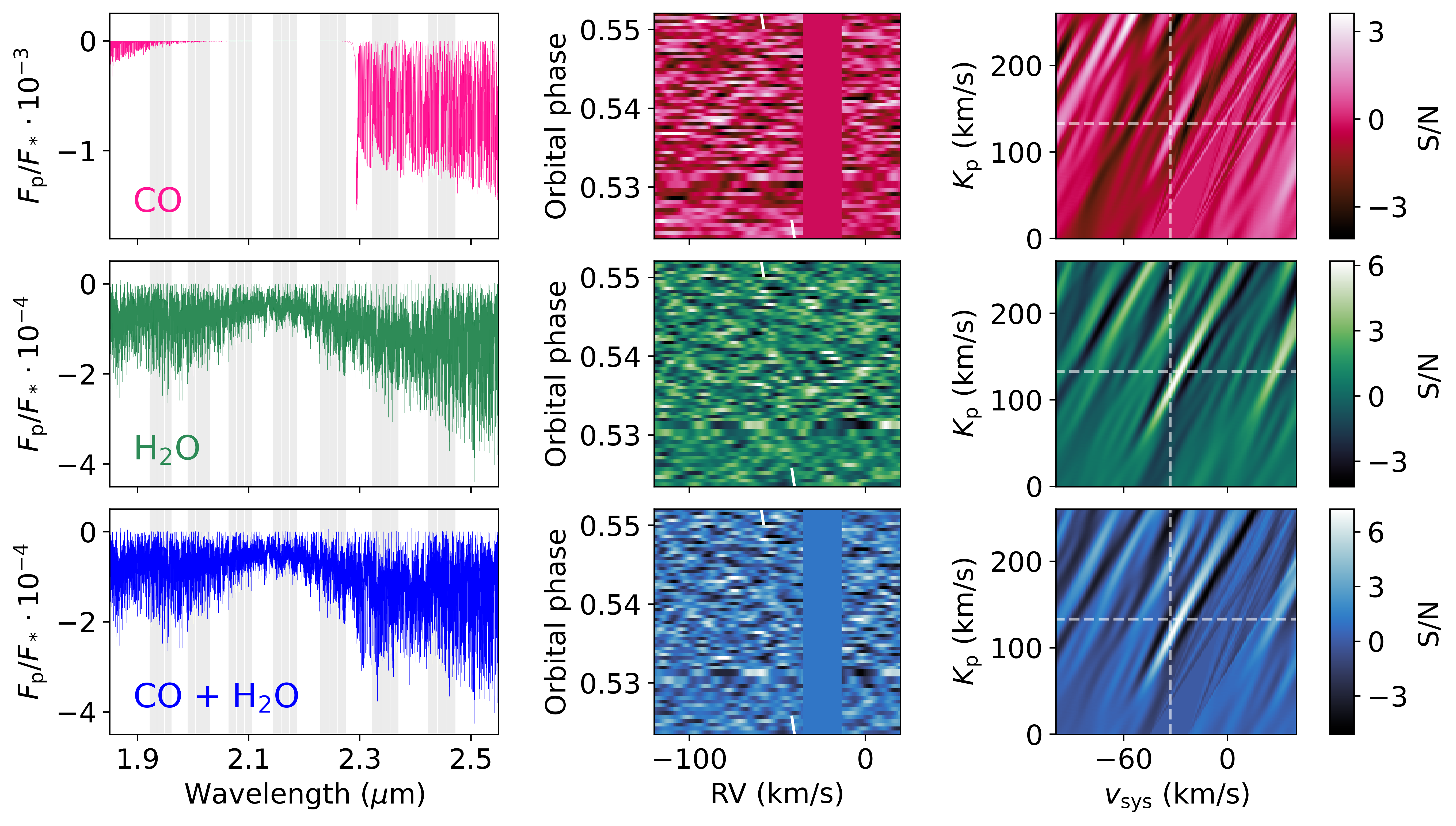}
        \caption{Model spectra, CCF maps, and S/N maps of CO, H$_2$O, and the combination of both species. {\it Left panels}: Model spectra normalized to the spectral continuum level. Wavelengths covered by the CRIRES$^+$ K2166 setting used in this work correspond to the gray shaded area. {\it Middle panels}: Cross-correlation function maps. The signal of H$_2$O and the signal obtained from a combination of CO and H$_2$O spectral lines can be identified as diagonal trails in the mappings, indicated by the solid white lines. A residual CO signal, which is consistent with the stellar rest frame and could not be entirely removed with \texttt{SYSREM}, is masked. {\it Right panels}: Signal-to-noise maps. These maps reveal no clear CO signal but a detection of spectral H$_2$O lines. The position of the planetary signal ($K_\mathrm{p}$ value reported by \cite{Brogi2013} and \cite{Birkby2017}; $\varv_\mathrm{sys}$ determined via the RV analysis in Sect.~\ref{sec:updated_ephemeris_from_RV_analysis}) is indicated by the dashed white lines. We show the CCF maps and S/N maps obtained after six consecutive {\tt SYSREM} iterations. A non-masked version of the CO and the species-combined CCF and S/N maps is given in Fig.~\ref{fig:SN-CCF-model-nomask}.}
        \label{fig:SN-CCF-model}
\end{figure*}

Each spectral model underwent two normalization steps. First, the models were divided by the blackbody spectrum of the host star. Then, the resulting planet-to-star flux ratio was normalized to the planetary continuum. Convolving the spectra with the instrumental profile yielded the final model spectra used for cross-correlation. The final model spectra used in this study are shown in the left panels of Fig.~\ref{fig:SN-CCF-model}.

We computed the weighted cross-correlation function (CCF) between each model spectrum and the residual spectra. For the $i$-th residual spectrum of the time series, the CCF is defined as
\begin{equation}
    \mathrm{CCF}_i\left(\varv\right) = \sum_j \frac{ R_{ij} m_j\left(\varv\right) }{ {\sigma_{ij}}^2 },
\end{equation}
with $R_{ij}$ being the residual spectral matrix, $\sigma_{ij}$ the corresponding uncertainties, and $m_j$ the spectral model shifted by a Doppler velocity in the range \mbox{--800\,km\,s$^{-1}$\,$ < \varv < $\,+800\,km\,s$^{-1}$}. For each wavelength segment, the CCFs of the spectral time series were arranged in a two-dimensional array. These arrays were then co-added to create a CCF map. Finally, the CCF maps from the two A and B nodding positions were combined into one final CCF map for each model spectrum. The middle panels of Fig.~\ref{fig:SN-CCF-model} illustrate the final CCF maps obtained using these procedures.

We aligned the CCF map to the rest frame of \FiftyOnePegb over a range of different orbital semi-amplitude velocity ($K_\mathrm{p}$) values. Under the assumption of a circular orbit, the planetary Doppler velocity is described by
\begin{equation}
    \label{equation:planetary-rest-frame} 
    \varv_\mathrm{p} \left( t \right) = \varv_\mathrm{sys} + \varv_\mathrm{bary} \left( t \right) + K_\mathrm{p} \sin{2\pi\phi \left( t \right)},
\end{equation}
where $\varv_\mathrm{sys}$ is the systemic velocity, $\varv_\mathrm{bary} \left( t \right)$ is the barycentric velocity correction, and $\phi \left( t \right)$ is the orbital phase. Each shifted CCF map was collapsed into a one-dimensional CCF by averaging along the time axis. The one-dimensional CCF of each alignment was further stacked into a two-dimensional array and normalized by its standard deviation, yielding the so-called S/N map. Following \cite{Cont2024, Cont2025}, the standard deviation was obtained by fitting the distribution of all CCF values with a Gaussian function. The S/N maps obtained in this work are shown in the right panels of Fig.~\ref{fig:SN-CCF-model}.

\subsection{Retrieval}
\label{sec:retrieval}

The procedures for retrieving the atmospheric properties of \FiftyOnePegb follow a framework similar to that described in Sect.~5.1.1 of \cite{Cont2024}. This framework relies on {\tt petitRADTRANS} to forward model the planetary emission spectrum \citep{Molliere2019}. 

To model the atmospheric thermal conditions, we used the \mbox{$T$-$p$} profile parameterization introduced by \cite{Pelletier2021}. Our choice is motivated by the fact that this parameterization does not make any a priori assumptions about the overall shape of the atmospheric thermal structure. Consequently, it can adapt flexibly to a wide range of possible $T$-$p$ profiles. We used six free temperature points ($T_1$--$T_6$), uniformly distributed in the logarithmic pressure space between $10^{-8}$\,bar and $10^2$\,bar. Following the methodology of \cite{Pelletier2021}, we adopted a smoothing prior on the $T$-$p$ profile with a standard deviation of $\sigma_\mathrm{s}$\,=\,150\,K\,dex$^{-2}$ to prevent nonphysical temperature oscillations in the retrieved temperature curve. Since the atmosphere of \FiftyOnePegb is predicted to host magnesium-silicate and iron clouds \citep{Goukenleuque2000}, we included a parameterization of an opaque cloud deck in our retrieval framework, which mutes the spectral contribution from atmospheric layers at pressures higher than $p_\mathrm{cloud}$.

The VMRs of the chemical species in \FiftyOnePegb's atmosphere were modeled with the {\tt FastChem} chemical equilibrium code as a function of the metallicity ([M/H]) and the C/O ratio \citep{Stock2022}\footnote{\url{https://github.com/exoclime/FastChem}}. We included the spectral contribution of CO and H$_2$O into the retrieval. These are the chemical species with the most dominant spectral lines in the K band. We also accounted for Rayleigh scattering from H$_2$ and He, as well as \mbox{H$_2$--H$_2$} and \mbox{H$_2$--He} collision-induced absorption effects.

The {\tt petitRADTRANS} code computes the flux per unit surface area and wavelength emerging from a planet's atmosphere. Therefore, the overall flux of \FiftyOnePegb is obtained via integration over the entire visible hemisphere, which requires knowledge of the planetary radius ($R_\mathrm{p}$). However, $R_\mathrm{p}$ of \FiftyOnePegb is poorly constrained but is expected to lie within the range \mbox{1.0--1.3\,$R_\mathrm{Jup}$} \citep{Birkby2017}. Consequently, we treated $R_\mathrm{p}$ as a free parameter and imposed an informed Gaussian prior centered on 1.15\,$R_\mathrm{Jup}$ with a standard deviation of 0.15\,$R_\mathrm{Jup}$. 

The surface gravity value ($\log{g}$) of \FiftyOnePegb is also unknown. This value is a function of $R_\mathrm{p}$ and a required input parameter of {\tt petitRADTRANS}. For this reason, we recalculated $\log{g}$ via $g = G M_\mathrm{p} / R_\mathrm{p}^2$ at each retrieval step with the updated $R_\mathrm{p}$ value. The planetary mass ($M_\mathrm{p}$) is calculated using $M_\mathrm{p}/M_* = K_*/K_\mathrm{p}$ and $G$ denotes the gravitational constant. The $K_\mathrm{p}$ value is updated at each retrieval step, and $M_*$ and $K_*$ denote the stellar mass and orbital semi-amplitude. Thus, in addition to the free parameters of our retrieval, we can estimate \FiftyOnePegb's $\log{g}$ and $M_\mathrm{p}$ values.

We also accounted for spectral line broadening due to the planetary rotation. To this end, we convolved the spectral forward model with a broadening profile. The broadening profile is parametrized by the projected equatorial rotation velocity ($\varv_\mathrm{rot} \sin{i}$). The analytical description of the broadening profile is given in Eq.~(3) of \cite{Diaz2011} and has been used successfully in previous studies \citep[e.g.,][]{Yan2023, Cont2024, Cont2025, Lesjak2025}.

\begin{figure*}
        \centering
        \includegraphics[width=\textwidth]{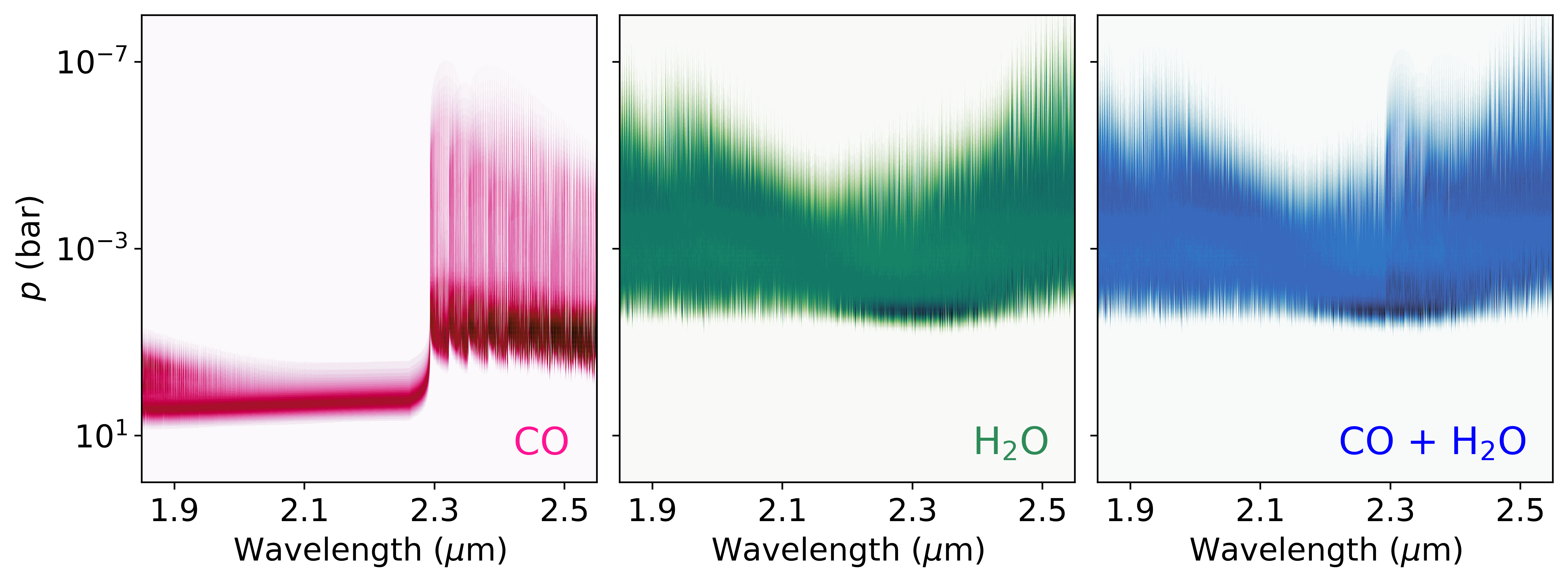}
        \caption{Contribution functions of the CO, H$_2$O, and species-combined model spectra. The CO signal originates from deeper regions of the atmosphere, whereas the H$_2$O lines form at higher altitudes. Consequently, in the species-combined model, the CO spectral lines are largely muted by the overlying H$_2$O signal. The contribution of CO to the combined signal becomes most evident in the region around the CO bandhead at approximately 2.3\,$\mu$m.}
        \label{fig:contribution_functions}
\end{figure*}

As described in Sect.~\ref{sec:cross_correlation}, the forward modeled spectrum was converted to the planet-to-star flux ratio and convolved with the CRIRES$^+$ instrument profile. Subsequently, for each individual spectrum of our time series, we used $K_\mathrm{p}$ and $\varv_\mathrm{sys}$ to Doppler shift the spectral forward model to the planetary rest frame. The shifted model spectrum was then interpolated to the wavelength solution of our observational data. These procedures resulted in a two-dimensional forward model matrix, with the same shape as the residual spectral matrix. We applied the filtering method of \cite{Gibson2022} to account for the presence of potential distortion effects introduced into our data by \texttt{SYSREM}.

We calculated the Gaussian log likelihood function
\begin{equation}
    \ln{L} = -\frac{1}{2}\sum_{i,j} \left[ \frac{\left(R_{ij} - M_{ij}\right)^2}{\left(\beta \sigma_{ij}\right)^2} + \ln{2 \pi \left(\beta \sigma_{ij}\right)^2} \right]
\end{equation}
to compare the model spectrum to the data. In this expression, $R_{ij}$ and $M_{ij}$ denote the individual elements of the residual spectral matrix and the filtered forward model matrix, respectively. The uncertainties of the residual spectra are represented by $\sigma_{ij}$ and $\beta$ acts as a noise scaling factor. We used the residual spectral matrix obtained after six consecutive {\tt SYSREM} iterations, at which $\Delta \sigma$ reached a plateau (see Sect.~\ref{sec:pre_processing_the_spectra}). For each wavelength segment and nodding position, the log likelihood function was calculated independently, and the resulting values were added to obtain the combined log likelihood function of all data. We estimated the model parameters by evaluating the combined log likelihood function using MCMC sampling with the {\tt emcee} package \citep{ForemanMackey2013}.

In summary, our high-resolution retrieval framework includes the following free parameters: the $T$-$p$ profile parameters $T_1$--$T_6$; the pressure level of an opaque cloud deck $p_\mathrm{cloud}$; the atmospheric chemical properties represented by [M/H] and the C/O ratio; the planetary radius $R_\mathrm{p}$; the projected equatorial rotation velocity $\varv_\mathrm{rot} \sin{i}$; the velocity parameters $K_\mathrm{p}$ and $\varv_\mathrm{sys}$; and the noise scaling parameter $\beta$. From these free parameters, additional planetary properties can be derived: the surface gravity $\log{g}$ and the planetary mass $M_\mathrm{p}$. For each free parameter, 32 walkers with 15\,000 steps were used in the sampling.

\section{Results and discussion}
\label{sec:results_and_discussion}

\subsection{Detection of the planetary spectral lines}
\label{sec:detection_of_the_planetary_spectral_lines}

We detect line features in the emission spectrum of \FiftyOnePegb's atmosphere. Figure~\ref{fig:SN-CCF-model} shows the S/N maps obtained by cross-correlating our CRIRES$^+$ spectra with the spectral models of CO, H$_2$O, and a combination of both species. The white dashed lines indicate the $K_\mathrm{p}$ reported by \cite{Brogi2013} and \cite{Birkby2017}, and the $\varv_\mathrm{sys}$ from our RV analysis in Sect.~\ref{sec:updated_ephemeris_from_RV_analysis}. The maps obtained under the inclusion of H$_2$O spectral lines show significant emission signals that coincide in velocity space but differ from the literature values by a few km\,s$^{-1}$. We provide a detailed discussion of this offset in Sect.~\ref{sec:retrieval_of_the_atmospheric_properties}, together with the results of our Bayesian retrieval analysis.

We identify a clear S/N peak of atmospheric H$_2$O with a detection strength of S/N\,=\,6.2. In contrast, we do not detect CO at a statistically significant level, with a S/N of only 1.9 at the expected planetary position. This is well below the S/N threshold of 4 commonly adopted in the literature. Nevertheless, we note that a diagonal feature appears at the exact \mbox{$K_\mathrm{p}$--$\varv_\mathrm{sys}$} position of the robust H$_2$O detection. While this weak CO feature alone cannot be considered statistically significant given the noise structure in the CO S/N map, its spatial coincidence with the H$_2$O signal is noteworthy, particularly in light of the previous CO detection reported by \cite{Brogi2013} using the original CRIRES instrument. A direct comparison with their result is not possible, as \cite{Brogi2013} do not provide a CO S/N map but only the combined CO + H$_2$O S/N map. Given that CRIRES$^+$ covers a broader wavelength range containing CO lines than the pre-upgraded CRIRES, one might expect a stronger CO signal in our data, which makes the absence of a significant detection in our analysis somewhat puzzling. However, the cumulative exposure time of our observations is considerably shorter ($\sim$3\,h) compared to that of \cite{Brogi2013} ($\sim$16\,h), which likely explains why the original CRIRES data achieved a significant CO detection despite its more limited spectral coverage. We note that a residual stellar CO signal is present in the data and could, in principle, bias the interpretation of the planetary signal if they overlap in the CCF map. As shown in Fig.~\ref{fig:SN-CCF-model}, this is not the case in our analysis, as the planetary signal is clearly separated from the stellar CO residuals in velocity space. Consequently, the planetary signal and our derived results remain unaffected by these residual features. Furthermore, the inclusion of CO into the retrieval in Sect.~\ref{sec:retrieval_of_the_atmospheric_properties} allowed us to place constraints on the C/O ratio, which would not be possible if the CO signal were purely spurious. Therefore, we consider it plausible that the diagonal CO feature with a S/N of 1.9 indicates a weak contribution of CO to the planetary spectrum. However, we do not claim a CO detection, and additional observations are required.

Cross-correlation with a model spectrum that includes both CO and H$_2$O spectral lines yields a detection peak with S/N\,=\,7.2. Comparison of this detection strength with that of the individual species reveals that the spectral signature of H$_2$O dominates the species-combined signal. This is also visible in the model spectra, where the combined spectrum of all species closely matches that of H$_2$O. The relatively small contribution of CO to the species-combined spectrum can be explained by the different atmospheric layers of \FiftyOnePegb in which CO and H$_2$O are spectroscopically active. The CO lines mostly form at higher pressures and deeper in the atmosphere, while the H$_2$O lines form at lower pressures and higher altitudes, as shown in Fig.~\ref{fig:contribution_functions}. Consequently, the CO spectral signature is largely obscured by the overlying H$_2$O spectral lines.

With an effective temperature of $5761\pm95$\,K \citep{Rosenthal2021}, 51\,Peg is expected to exhibit an elevated number of strong CO lines in its stellar spectrum. We observed that residual stellar CO features persisted in our data throughout the entire range of {\tt SYSREM} iterations. Therefore, we masked the CO and species-combined CCF maps within a $\varv_\mathrm{sys}$ interval of $\pm 10$\,km\,s$^{-1}$ around the stellar RV. This procedure propagated into the respective S/N maps as a small masked region. However, the affected region does not overlap with the planetary signal of \FiftyOnePegb, and thus it does not affect our conclusion that a robust signal is found in the \mbox{CO + H$_2$O} combined analysis and that a minor CO feature at the same position as the detected H$_2$O signal can be identified. An unmasked version of the CCF and S/N maps is provided in Fig.~\ref{fig:SN-CCF-model-nomask}.

The detection of absorption lines in the dayside emission spectrum of \FiftyOnePegb provides strong evidence of the presence of a non-inverted $T$-$p$ profile. This finding is in line with previous investigations of other hot Jupiter atmospheres \citep[e.g.,][]{Bazinet2024, Finnerty2024}.

\subsection{Retrieval of atmospheric and planetary properties}
\label{sec:retrieval_of_the_atmospheric_properties}

Our retrieval is able to constrain the atmospheric thermal and chemical structure, and it allowed us to derive additional planetary properties from the Doppler shift of \FiftyOnePegb's orbital motion. Figure~\ref{fig:corner-plot} shows the posterior distributions, together with the correlations between the model parameters. A summary of the best-fit parameters of the retrieval is provided in Table~\ref{table:retrieval-results}.

\subsubsection{Free parameters}
\label{sec:free_parameters}

\begin{figure*}
        \centering
        \includegraphics[width=\textwidth]{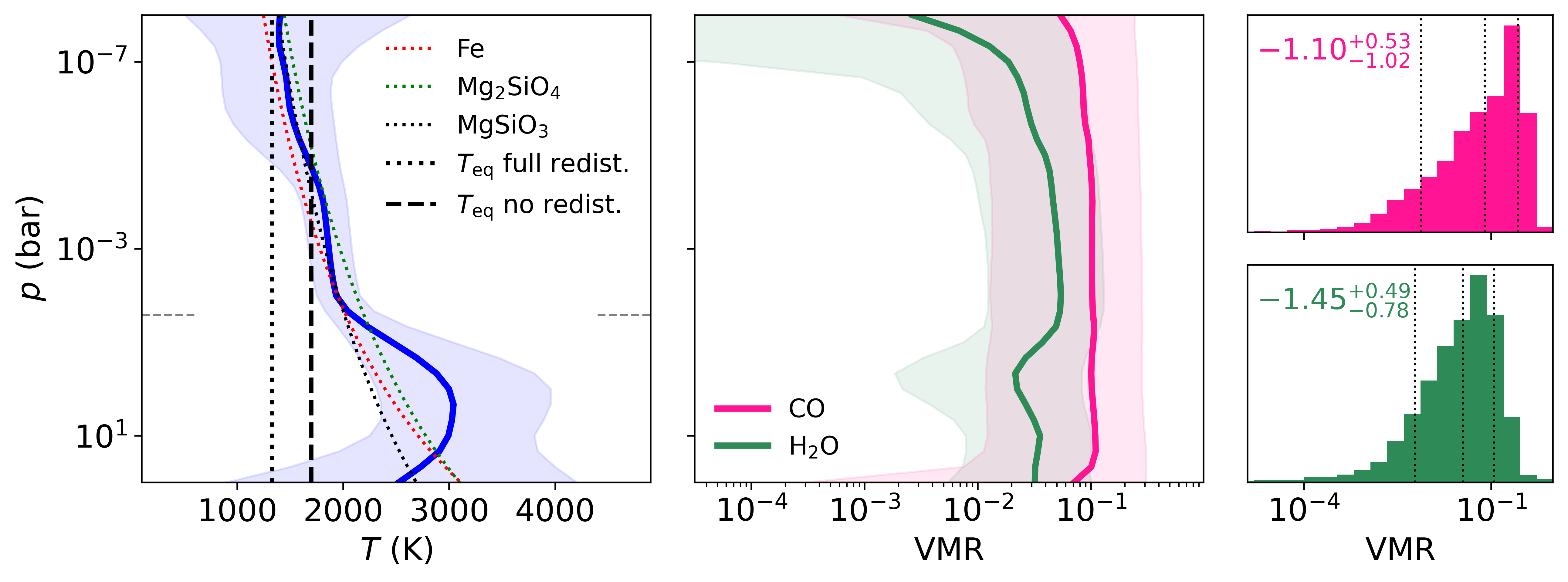}
        \caption{Atmospheric temperature and molecular abundance profiles. \textit{Left panel}: Median $T$-$p$ profile (solid line) with its one-sigma uncertainty range (shaded area). The vertical dotted and dashed lines indicate the expected $T_\mathrm{eq}$ of \FiftyOnePegb, calculated for the two limiting cases of full and no dayside to nightside heat redistribution, respectively. The horizontal gray lines at $10^{-1.6}$\,bar mark the retrieved lower limit on $p_\mathrm{cloud}$. In addition, we plot the condensation curves of Fe, Mg$_2$SiO$_4$, and MgSiO$_3$, the most important cloud constituents expected in \FiftyOnePegb's atmosphere. \textit{Middle panel}: Retrieved molecular VMRs (solid lines) together with their one-sigma uncertainties (shaded areas). Both CO and H$_2$O remain approximately constant throughout the entire pressure range. \textit{Right panels}: Histograms of the VMRs of the two molecular species, sampled at a representative pressure of $10^{-4}$\,bar. The median and one-sigma uncertainty intervals are indicated by the dotted black lines.}
        \label{fig:Tp-VMRs}
\end{figure*}

Figure~\ref{fig:Tp-VMRs} shows the retrieved $T$-$p$ profile, which confirms the presence of a non-inverted thermal structure already detected in Sect.~\ref{sec:detection_of_the_planetary_spectral_lines}. This finding is consistent with the thermal properties derived in previous observational studies of \FiftyOnePegb \citep{Brogi2013, Birkby2017, Chiavassa2019}. Our retrieval yields a $T$-$p$ profile spanning between approximately 1400\,K to 3000\,K from the upper to the lower planetary atmosphere. The retrieved thermal conditions of \FiftyOnePegb's upper atmosphere are consistent with the planetary $T_\mathrm{eq}$, estimated via Eq.~(4) of \cite{CowanAgol2011} as approximately 1330\,K and 1700\,K in the extreme cases of full and no heat redistribution from the planet's dayside to nightside. Toward the bottom of \FiftyOnePegb's atmosphere the retrieved $T$-$p$ profile is hotter than the calculated $T_\mathrm{eq}$ values. This is in line with the expectation that hot Jupiters cannot efficiently reradiate significant amounts of energy received from their host stars back to space from their lower atmospheres \citep[e.g.,][]{Fortney2008, Guillot2010}. 

Our retrieval constrains the pressure level below which an opaque cloud deck may exist to $p_\mathrm{cloud} > 10^{-1.6}$\,bar. In Figure~\ref{fig:Tp-VMRs}, we overplot the condensation curves of Fe, Mg$_2$SiO$_4$, and MgSiO$_3$ from \cite{Visscher2010} on the retrieved $T$-$p$ profile. In the deeper layers of \FiftyOnePegb’s atmosphere temperatures are too high for these species to condense. The condensation curves intersect the retrieved $T$-$p$ profile at approximately the same pressure as our derived $p_\mathrm{cloud}$ limit. At low atmospheric pressures, the condensation curves remain close to the $T$-$p$ profile, suggesting that condensation could in principle also occur higher up in the atmosphere. However, the retrieval excludes the presence of a cloud deck in this low-pressure region. Consequently, the most likely cloud formation region lies near a pressure of $\sim$10$^{-1.6}$\,bar, where the condensation curves first intersect the $T$-$p$ profile and where the retrieval allows for the presence of a cloud deck. The $\log{p_\mathrm{cloud}}$ posterior distribution is very similar to those obtained from high-resolution spectroscopy studies of other hot Jupiters with comparable equilibrium temperatures and surface gravities, such as WASP-43b \citep{Lesjak2023} and HD~189733b \citep{Boucher2021, Finnerty2024}, noting that we report a two-sigma lower limit rather than median values. In contrast, the cloud deck pressure retrieved for $\tau$\,Boo\,b by \cite{Pelletier2021} is higher, consistent with the higher $\log{g}$ of that planet.

The C/O ratio in \FiftyOnePegb's atmosphere is constrained to $0.54_{-0.18}^{+0.23}$, while we derived [M/H]\,=\,$2.63_{-1.00}^{+0.93}$\,dex relative to the solar value. The derived C/O ratio of \FiftyOnePegb is consistent with that of its host star, estimated as approximately 0.5 and 0.7 when using the abundance values of \cite{Brewer2016} and \cite{Maldonado2018}, respectively. The planetary [M/H] value is significantly higher than that of the host star, reported as 0.2\,dex in both studies. The retrieved value for [M/H] is high even when compared to the Saturn mass planet HD~149026b which is considered the most metal-rich giant planet atmosphere detected to date \citep{Bean2023}. However, \FiftyOnePegb’s mass lies at the border to the hot Saturn mass regime, where planet formation and interior models predict substantially enhanced heavy-element enrichment. This is also reflected in the observed relation between planet mass and atmospheric (and bulk) metallicity \citep{Kreidberg14, Welbanks2019, Sun2024}. It is possible that the high atmospheric metallicity measured for \FiftyOnePegb in this study could be a reflection of a quenched atmosphere dominated by heavy elements from the planet interior. In addition, we note that our retrieved [M/H] uncertainties of approximately $\pm1$\,dex are relatively large. Thus, lower metallicity values at the 1.5\,dex level, which have previously been found in other gas giant exoplanets, are consistent with our measurement.

Figure~\ref{fig:Tp-VMRs} shows how the retrieved [M/H] and C/O values translate into the VMRs of CO and H$_2$O as a function of atmospheric pressure. The VMRs are approximately constant throughout the investigated pressure range, except for H$_2$O at the lowest pressures, where a slight decrease can be observed. At a representative atmospheric pressure of $10^{-4}$\,bar, the retrieved logarithmic VMRs of CO and H$_2$O are $-1.10_{-1.02}^{+0.53}$ and $-1.45_{-0.78}^{+0.49}$, respectively, which is higher than previous estimates via cross-correlation \citep{Brogi2013, Birkby2017}. We note that the relatively constant VMR profiles of CO and H$_2$O hint -- as mentioned in the previous paragraph -- toward significant chemical quenching, i.e., atmospheric mixing timescales are shorter than chemical reaction timescales, causing the molecular abundances to remain fixed at the values set in the deeper atmosphere \citep{Moses2014, Heng2016}. However, our retrieval framework assumes thermochemical equilibrium and does not explicitly include quenching. We acknowledge that the assumption of thermochemical equilibrium represents a limitation of our retrieval, as the exclusion of quenching could potentially bias the inferred chemical abundances. This may explain the high [M/H] value retrieved, as a higher metallicity is required under the assumption of equilibrium chemistry to match the elevated molecular abundances resulting from quenching.

To ensure that the high [M/H] value is not the result of including CO emission lines, which contribute negligibly to the planetary detection in Sect.~\ref{sec:cross_correlation}, we performed an additional retrieval considering only H$_2$O and excluding CO (the C/O ratio is fixed to the stellar value in this retrieval to exclude any dependence on CO). This retrieval converged on the same parameter values as the retrieval that included CO. Therefore, we conclude that our results are driven by the strong H$_2$O signal and that the elevated [M/H] value is not a consequence of including CO in our analysis.

For \FiftyOnePegb's planetary radius, we applied a Gaussian prior based on the previous estimate of 1.01--1.32\,$R_\mathrm{Jup}$ from \cite{Birkby2013}. Our retrieval yields $R_\mathrm{p}=1.07\pm0.14\,R_\mathrm{Jup}$, which is in agreement with this earlier estimate.

We observe a spectral line width that exceeds the broadening expected from \FiftyOnePegb's planetary rotation under the assumption of tidal locking. The derived projected equatorial rotation velocity is $\varv_\mathrm{rot} \sin{i}$\,=\,$4.07_{-0.60}^{+0.53}$\,km\,s$^{-1}$, whereas the expected tidally locked value lies between 1.2\,km\,s$^{-1}$ and 1.6\,km\,s$^{-1}$ for $R_\mathrm{p}$ values between 1.0\,$R_\mathrm{Jup}$ and 1.3\,$R_\mathrm{Jup}$. The presence of a velocity excess in the retrieved $\varv_\mathrm{rot} \sin{i}$ value suggests strong atmospheric dynamics, as phenomena such as super-rotating winds or turbulent flow can produce additional line broadening beyond what is expected from planetary solid body rotation. Strong jet stream winds have already been observed in the atmospheres of other gas giant exoplanets, such as \mbox{WASP-127b} and \mbox{WASP-121b} \citep{Nortmann2025, Seidel2025}. Our interpretation that jet stream winds or turbulent mixing are responsible for the excess line broadening in \FiftyOnePegb's spectrum rather than a dayside to nightside flow regime is motivated by the fact that the observed signal is not affected by an offset in the retrieved $\varv_\mathrm{sys}$ \citep[e.g.,][]{AlonsoFloriano2019, SanchezLopez2019, Yan2022}. Indeed, \cite{Lesjak2025} have shown that a dayside to nightside wind flow would result in an asymmetric spectral line shape, producing a net offset from the expected position in the $\varv_\mathrm{sys}$ value.

We emphasize that care must be taken to ensure that the measured spectral line broadening is not biased by instrumental effects, data processing, or overly simplistic modeling. Several effects, such as overestimation of the instrumental spectral resolution, wavelength instabilities during the spectral time series, erosion of line shapes by {\tt SYSREM}, or neglecting  atmospheric pressure broadening, can introduce biases into the inferred rotational broadening if not properly accounted for. In our analysis, we have addressed these effects as follows: the spectral resolution was estimated via {\tt molecfit} fitting, which we consider a reliable method. Spectral drifts were corrected during data preprocessing steps. The potential deformation of line shapes by {\tt SYSREM} was accounted for by filtering our forward model spectra according to \cite{Gibson2022}. Finally, atmospheric pressure broadening is included in the spectral forward models via the opacities of the radiative transfer calculation. We therefore conclude that the observed line broadening is likely of atmospheric origin, consistent with findings from other hot gas giant exoplanets that also exhibit excess spectral line broadening \citep[e.g.,][]{Cont2022, Lesjak2025, Zhang2026}.

The noise scaling term $\beta$ is close to one. This indicates a proper estimate of the CRIRES$^+$ uncertainties.

\subsubsection{Deviation from previously reported $K_\mathrm{p}$ values}
\label{sec:Kpoffset}

The systemic velocity is derived as $\varv_\mathrm{sys} = -35.21_{-2.35}^{+2.11}$\,km\,s$^{-1}$, which is consistent with the value inferred from the stellar spectral lines in Sect.~\ref{sec:updated_ephemeris_from_RV_analysis}. The retrieved $K_\mathrm{p}$ is equal to $102.8_{-9.1}^{+8.3}$\,km\,s$^{-1}$. Previous cross-correlation studies of \FiftyOnePegb have reported higher $K_\mathrm{p}$ values than those obtained from our retrieval. \citet{Brogi2013} and \citet{Birkby2017} both found values around 133\,km\,s$^{-1}$, which is approximately 30\,km\,s$^{-1}$ higher than our results. 

We tested if adopting an orbital phase shift ($\Delta\phi$) could align the signal of \FiftyOnePegb to the $K_\mathrm{p}$ values reported by these previous studies. To this end, we conducted an additional retrieval run in which we fixed all parameters to their best-fit values and allowed only for an additional $\Delta \phi$ as a free parameter. This retrieval resulted in a best-fit phase shift of $\Delta\phi=-0.007$ for aligning our signal with a $K_\mathrm{p}$ of 133\,km\,s$^{-1}$. 
We verified whether the uncertainty on the orbital phase, resulting from the propagation of the uncertainties in the planetary time of conjunction ($T_\mathrm{C}$) and $P_\mathrm{orb}$, can allow for such a phase shift. The orbital phase uncertainty is $\pm0.0007$, which is an order of magnitude smaller than the phase shift value required to match the $K_\mathrm{p}$ from the previous studies. Therefore, we rule out the possibility that uncertainties in $T_\mathrm{C}$ and $P_\mathrm{orb}$ are the cause of the observed $K_\mathrm{p}$ discrepancy.

Alternatively, a nonzero eccentricity could cause an orbital phase shift. The observations of \cite{Brogi2013}, \cite{Birkby2017}, and the present study probe different orbital phase intervals. Consequently, an eccentric orbit would produce distinct phase shifts in each dataset, leading to different offsets in the derived $K_\mathrm{p}$ values. However, we consider it unlikely that a mature hot Jupiter such as \FiftyOnePegb has a significantly eccentric orbit. Indeed, in Sect.~\ref{sec:updated_ephemeris_from_RV_analysis}, we derived a two-sigma upper eccentricity limit at 0.0063, which is consistent with a circular orbit. However, we note that this does not rule out the very small eccentricity of 0.001 and the offset in the argument of periastron of 6\,deg that would align the signal of \cite{Birkby2017} with the expected $\varv_\mathrm{sys}$ value.

Direct comparison with the results of \cite{Brogi2013} and \citet{Birkby2017} is complicated by the phase shifts they applied to align their planetary signals with the expected $\varv_\mathrm{sys}$. We note that the phase shift of 0.1 in \cite{Birkby2017} is erroneously expressed in radians rather than in orbital phase units (confirmed by the authors; priv. comm.), corresponding to an orbital phase shift of 0.016, which is close to the 0.0095 phase shift applied by \cite{Brogi2013}. Both previous studies find higher $K_\mathrm{p}$ values than retrieved in our work. Considering the results for individual nights in \cite{Brogi2013} (see their Table~1; $K_\mathrm{p}$ values of $122.3_{-24.8}^{+15.9}$\,km\,s$^{-1}$ and $134.7_{-21.3}^{+13.7}$\,km\,s$^{-1}$) and the one-sigma uncertainty interval of \cite{Birkby2017} (see their Fig.~8, with the one-sigma region extending down to a $K_\mathrm{p}$ of approximately 107\,km\,s$^{-1}$), our result of \mbox{$K_\mathrm{p}=102.8_{-9.1}^{+8.3}$\,km\,s$^{-1}$} is close to agreement at the upper uncertainty margin. However, our $K_\mathrm{p}$ cannot be fully reconciled with the combined result of all nights from \cite{Brogi2013}. Part of this discrepancy may stem from the different orbital phases sampled: \cite{Brogi2013} observed closer to the extrema of \FiftyOnePegb's RV curve, where the sensitivity to $K_\mathrm{p}$ is highest, whereas our observations near superior conjunction are less sensitive to the parameter. Additionally, discrepancies between the studies may potentially arise from a different treatment of wavelength calibration and telluric removal methods, both of which are known caveats in high-resolution spectroscopy of exoplanet atmospheres. We therefore present our measurement as an alternative orbital solution, and note that resolving this discrepancy will require future observations.

To assess the robustness of our results, we performed an additional retrieval with $K_\mathrm{p}$ fixed to 133\,km\,s$^{-1}$, close to the value reported by \citet{Brogi2013} and \citet{Birkby2017}. The posterior distributions resulting from this additional test are shown in Fig.~\ref{fig:corner-plot-Kp133}. All retrieved parameters are consistent with those from our free $K_\mathrm{p}$ retrieval, except for $\varv_\mathrm{sys}$, which is red-shifted by approximately 6\,km\,s$^{-1}$ relative to the value derived in Sect.~\ref{sec:updated_ephemeris_from_RV_analysis} from the stellar spectral lines. Such an offset in $\varv_\mathrm{sys}$ could in principle result from a strong dayside to nightside wind. However, we deem this scenario unlikely, as dayside to nightside circulation is expected at $T_\mathrm{eq}$ values higher than that of \FiftyOnePegb, for which jet stream circulation is predicted \citep[e.g.,][]{Komacek2016, Showman2020}. Furthermore, the retrieved spectral line broadening is slightly larger in the fixed $K_\mathrm{p}$ retrieval than in the free $K_\mathrm{p}$ retrieval, suggesting that $K_\mathrm{p} \sim 103$\,km\,s$^{-1}$ achieves a better alignment of the planetary spectral lines to a common rest frame.

\begin{table}
        \caption{Results of atmospheric retrieval on \FiftyOnePegb.
        }
        \label{table:retrieval-results}
        \centering
        \renewcommand{\arraystretch}{1.5}
        \begin{threeparttable}
                \begin{tabular}{l c c c}
                        \hline\hline
                        \noalign{\smallskip}
                        Parameter & Value  & Uniform prior & Unit  \\
                        \noalign{\smallskip}
                        \hline
                        \noalign{\smallskip}
                        \multicolumn{4}{l}{\it Free parameters} \\
                        \noalign{\smallskip}
                        $T_1$                      & $< 3507$  & (100, 5000)   & K \\
                        $T_2$                      & $1496_{-609}^{+398}$  & (100, 5000)   & K \\
                        $T_3$                      & $1812_{-213}^{+222}$  & (100, 5000)   & K \\
                        $T_4$                      & $1932_{-184}^{+225}$  & (100, 5000)   & K \\
                        $T_5$                      & $2997_{-683}^{+960}$  & (100, 5000)   & K \\
                        $T_6$                      & no constraint                                  & (100, 5000)   & K \\
                        $\log{p_\mathrm{cloud}}$   & $ > -1.58 $                                    & (--8, 2)      & $\log{\mathrm{bar}}$ \\
                        $\mathrm{[M/H]}$           & $2.63_{-1.00}^{+0.93}$  & (--5, 5)       & dex \\
                        $\mathrm{C/O}$             & $0.54_{-0.23}^{+0.18}$  & (0, 1.5)      & \ldots \\
                        $R_\mathrm{p}$             & $1.07 \pm 0.14$  & (0.1, 3)      & $R_\mathrm{Jup}$ \\
                        $K_\mathrm{p}$             & $102.8_{-9.1}^{+8.3}$  & (0, 260)      & km\,s$^{-1}$ \\
                        $\varv_\mathrm{sys}$        & $-35.21_{-2.35}^{+2.11}$  & (--60, 0)     & km\,s$^{-1}$ \\
                        $\varv_\mathrm{rot}\sin{i}$ & $4.07_{-0.60}^{+0.53}$  & (0, 10)       & km\,s$^{-1}$ \\
                        $\beta$                     & $1.0756_{-0.0005}^{+0.0006}$   & (0, 3)         &    \ldots \\ 
                        \noalign{\smallskip}
                        \hline
                        \noalign{\smallskip}
                        \multicolumn{4}{l}{\it Derived parameters} \\
                        \noalign{\smallskip}
                        $\log{g}$              & $3.15 \pm 0.12$  & \ldots & $\log{\mathrm{cm\,s}^{-2}}$ \\
                        $M_\mathrm{p}$         & $0.61_{-0.05}^{+0.06}$  & \ldots & $M_\mathrm{Jup}$ \\
                        $i$                    & $49.8_{-5.7}^{+5.8}$  & \ldots & deg \\
                        \noalign{\smallskip}
                        \hline
                \end{tabular}
        \end{threeparttable}
\end{table}

\subsubsection{Derived parameters}
\label{sec:derived_parameters}

As described in Sect.~\ref{sec:retrieval}, the retrieval parameters $R_\mathrm{p}$ and $K_\mathrm{p}$ enable the derivation of quantitative constraints on additional planetary properties. These include the surface gravity $\log{g}$ and planetary mass $M_\mathrm{p}$. In addition, assuming the planet's rotation axis is aligned with the inclination angle of its orbit, the orbital inclination can be derived from the retrieved $K_\mathrm{p}$ value via $\sin{i} = P_\mathrm{orb} K_\mathrm{p} / 2 \pi a$, where $P_\mathrm{orb}$ is the orbital period and $a$ the orbital semi-major axis. Consequently, the orbital inclination~$i$ represents a third parameter of \FiftyOnePegb that can be constrained from the retrieval, in addition to $\log{g}$ and $M_\mathrm{p}$. Figure~\ref{fig:corner-plot-derived-parameters} shows the posterior distributions of the three derived parameters, all of which result in bounded constraints, yielding the following values: $\log{g}$\,=\,$3.15 \pm 0.12$\,$\log{\mathrm{cm\,s}^{-2}}$, $M_\mathrm{p}$\,=\,$0.61_{-0.05}^{+0.06}$\,$M_\mathrm{Jup}$, and $i$\,=\,$49.8_{-5.7}^{+5.8}$\,deg. For \FiftyOnePegb, this is the first direct measurement of $\log{g}$, with a value that is consistent with those typically found for other hot Jupiters \citep[e.g.,][]{Southworth2007, Torres2008}. The two parameters $M_\mathrm{p}$ and $i$ were previously estimated by \cite{Brogi2013} and \cite{Birkby2017}. The values of 0.46\,$M_\mathrm{Jup}$ and 0.48\,$M_\mathrm{Jup}$ in these studies differ from $M_\mathrm{p}$ determined in this work, due to the direct dependence of the parameter on $K_\mathrm{p}$. Our derived orbital inclination is lower than the previously reported values of approximately 80\,deg. Again, this discrepancy can be attributed to the dependence on $K_\mathrm{p}$. Our inferred value of $i = 50$\,deg is supported by the recent measurement of the stellar inclination of 51\,Peg of $53_{-11}^{+25}$\,deg \citep{Metcalfe2024}, under the assumption that the planetary orbit is aligned with the stellar spin axis. We note, however, that due to the large uncertainties, this measurement is also consistent with the inclination value reported by \citet{Birkby2017} and only marginally inconsistent with that of \citet{Brogi2013}.

The determination of the orbital inclination is needed for understanding which regions in the planetary atmosphere are being probed and provides context to the retrieved atmospheric thermal, chemical, and dynamical properties. While the results of \cite{Brogi2013} and \cite{Birkby2017} suggest that around superior conjunction the planetary dayside is almost perfectly aligned with the observer's line of sight, the lower inclination value derived in the present work indicates that a substantial fraction of the planetary nightside may still be visible during these orbital phases. Therefore, the atmospheric conditions obtained from our retrieval are likely not corresponding to pure dayside conditions, but rather to a combination of dayside and nightside conditions present on the hemisphere aligned with the observer's line of sight. To resolve the discrepancy between previous studies and the present work in the derived $M_\mathrm{p}$ and $i$ values, and to enable a more comprehensive understanding of \FiftyOnePegb's atmospheric properties, future observations covering a broader range of orbital phases will be required.

The derivation of $K_\mathrm{p}$ provides an interesting additional constraint on the stellar mass that is implicitly included in the calculation of the orbital parameters. We can write it in an explicit way if we use Kepler's third law to derive mass constraints that depend only on the orbital period $P_\mathrm{orb}$, eccentricity $e$, and the semi-amplitude velocities $K_\mathrm{p}$ and $K_*$:

\begin{eqnarray}
        M_\mathrm{p} \left(\sin{i} \right)^3 & = & \frac{P_\mathrm{orb}}{2 \pi G} \left( 1 - e^2\right)^{\frac{3}{2}} \frac{\left( K_* + K_\mathrm{p} \right)^3}{1 + \frac{K_\mathrm{p}}{K_*}},\\
        M_* \left(\sin{i} \right)^3 & = & \frac{P_\mathrm{orb}}{2 \pi G} \left( 1 - e^2\right)^{\frac{3}{2}} \frac{\left( K_* + K_\mathrm{p} \right)^3}{1 + \frac{K_*}{K_\mathrm{p}}}.
\end{eqnarray}

From our results, we obtained $M_\mathrm{p} (\sin{i})^3 = 0.27$\,$M_{\rm Jup}$ and $M_* (\sin{i})^3 = 0.48$\,$M_\odot$. This is consistent with the mass determined from spectroscopic analysis and isochrone models, \mbox{$M_*$\,=\,1.07\,$M_\odot$}, which also yields $i = 50$\,deg and $M_\mathrm{p} = 0.61$\,$M_{\rm Jup}$.

%

\section{Conclusions}
\label{sec:conclusions}

We have characterized the atmospheric properties and derived planetary parameters of the non-transiting hot Jupiter \FiftyOnePegb using high-resolution spectroscopy in the K band with CRIRES$^+$. In addition, we have used archival data from multiple high-stability, high-resolution spectrographs to refine the planetary ephemeris. By combining RV analysis methods, the cross-correlation technique, and a Bayesian retrieval framework, we were able to obtain the following key results:

\begin{enumerate}

\item Our RV analysis provides the most up-to-date estimates of the time of conjunction and orbital period of \FiftyOnePegb. Moreover, we derived an upper limit on the orbital eccentricity that is consistent with a circular orbit (Sect.~\ref{sec:updated_ephemeris_from_RV_analysis}). The retrieved orbital semi-amplitude of \FiftyOnePegb is lower than reported in previous cross-correlation studies (Sect.~\ref{sec:Kpoffset});

\item We confirmed the spectral signature of H$_2$O and found a possible indication for CO spectral lines in the planetary emission spectrum. The CO signal appears weaker because it originates from deeper atmospheric layers than the H$_2$O spectral lines and is therefore largely muted (Sect.~\ref{sec:detection_of_the_planetary_spectral_lines});

\item We inferred a non-inverted $T$-$p$ profile, consistent with previous studies of \FiftyOnePegb and with the expected equilibrium temperature. This indicates the absence of strong optical absorbers capable of producing thermal inversions. The retrieval constrains the pressure level of an opaque cloud deck at the bottom of \FiftyOnePegb's atmosphere, which is consistent with the condensation curves of magnesium-silicate and iron clouds (Sect.~\ref{sec:free_parameters});

\item Our retrieval yields a C/O ratio consistent with the solar value. The atmospheric metallicity retrieved under the assumption of chemical equilibrium is higher than typically expected for hot Jupiters. This elevated value may result from a quenched chemical structure of \FiftyOnePegb's atmosphere, which is caused by the uplift of heavy elements from the planet's interior (Sect.~\ref{sec:free_parameters});

\item We inferred the presence of spectral line broadening that cannot be explained by \FiftyOnePegb's tidally locked rotation alone. This excess broadening is most likely caused by super-rotation in the planetary atmosphere. The absence of an offset in the retrieved systemic velocity from the expected value disfavors a dayside to nightside wind flow as the main cause of the observed line broadening (Sect.~\ref{sec:free_parameters});

\item Finally, we derived additional properties of \FiftyOnePegb from the retrieved RV semi-amplitude, including the planetary surface gravity, mass, and orbital inclination. Given that our RV semi-amplitude differs from previous estimates, the derived mass is higher and the orbital inclination lower than reported in earlier studies (Sect.~\ref{sec:derived_parameters}).

\end{enumerate}

Beyond analyzing the atmosphere of \FiftyOnePegb, the present study demonstrates that high-resolution spectroscopy is a powerful tool for investigating fundamental parameters  -- such as planetary mass or orbital inclination -- that are otherwise difficult to access in non-transiting exoplanet systems. Several massive, short-period exoplanets that are not observable in transit still lack measurements of these parameters \citep[e.g.,][]{Cullen2024}. Therefore, future high-resolution emission spectroscopy of non-transiting exoplanets will enable a significant expansion of the sample of planets with well constrained fundamental properties, as well as provide detailed atmospheric characterization.

%

\section*{Data availability}
The extracted data are available at \url{https://zenodo.org/records/17565920} \citep{cont_2025_17565920}.

\begin{acknowledgements}
    CRIRES$^+$ is an ESO upgrade project carried out by Th\"{u}ringer Landessternwarte Tautenburg, Georg-August Universit\"{a}t G\"{o}ttingen, and Uppsala University. The project is funded by the Federal Ministry of Education and Research (Germany) through Grants 05A11MG3, 05A14MG4, 05A17MG2 and the Knut and Alice Wallenberg Foundation. This project is based on observations collected at the European Organisation for Astronomical Research in the Southern Hemisphere under the ESO programme 114.279X. 
    D.C. is supported by the LMU-Munich Fraunhofer-Schwarzschild Fellowship and by the Deutsche Forschungsgemeinschaft (DFG, German Research Foundation) under Germany's Excellence Strategy -- EXC 2094 -- 390783311. 
    F.L. acknowledges support from the European Union (ERC-CoG, EVAPORATOR, Grant agreement No. 101170037). Views and opinions expressed are however those of the author(s) only and do not necessarily reflect those of the European Union or the European Research Council. Neither the European Union nor the granting authority can be held responsible for them. 
    D.S. acknowledges financial support from the project PID2021-126365NB-C21(MCI/AEI/FEDER, UE) and from the Severo Ochoa grant CEX2021-001131-S funded by MCIN/AEI/ 10.13039/501100011033. 
    N.P., L.B.-Ch., and A.D.R. acknowledge support by the Knut and Alice Wallenberg Foundation (grant 2018.0192). 
    O.K. acknowledges support by the Swedish Research Council (grant agreements no. 2023-03667) and by the Swedish National Space Agency. 
    E.N. acknowledges the support from the Deutsches Zentrum für Luft- und Raumfahrt (DLR, German Aerospace Center) -- project number 50OP2502. 
    M.R. acknowledges the support by the DFG priority program SPP 1992 ``Exploring the Diversity of Extrasolar Planets'' (DFG PR 36 24602/41). 
    This work has made use of the following Python packages: {\tt Astropy} \citep{AstropyCollaboration2013}, {\tt CMasher} \citep{vanderVelden2020}, {\tt corner} \citep{ForemanMackey2013}, {\tt Matplotlib} \citep{Hunter2007}, {\tt NumPy} \citep{Harris2020}, {\tt PyAstronomy} \citep{Czesla2019}, and {\tt SciPy} \citep{Virtanen2020}.
\end{acknowledgements}


\bibliographystyle{aa} 
\bibliography{references}


\appendix

\FloatBarrier

\section{Additional figures}

\begin{figure}[h]
        \centering
        \includegraphics[width=\columnwidth]{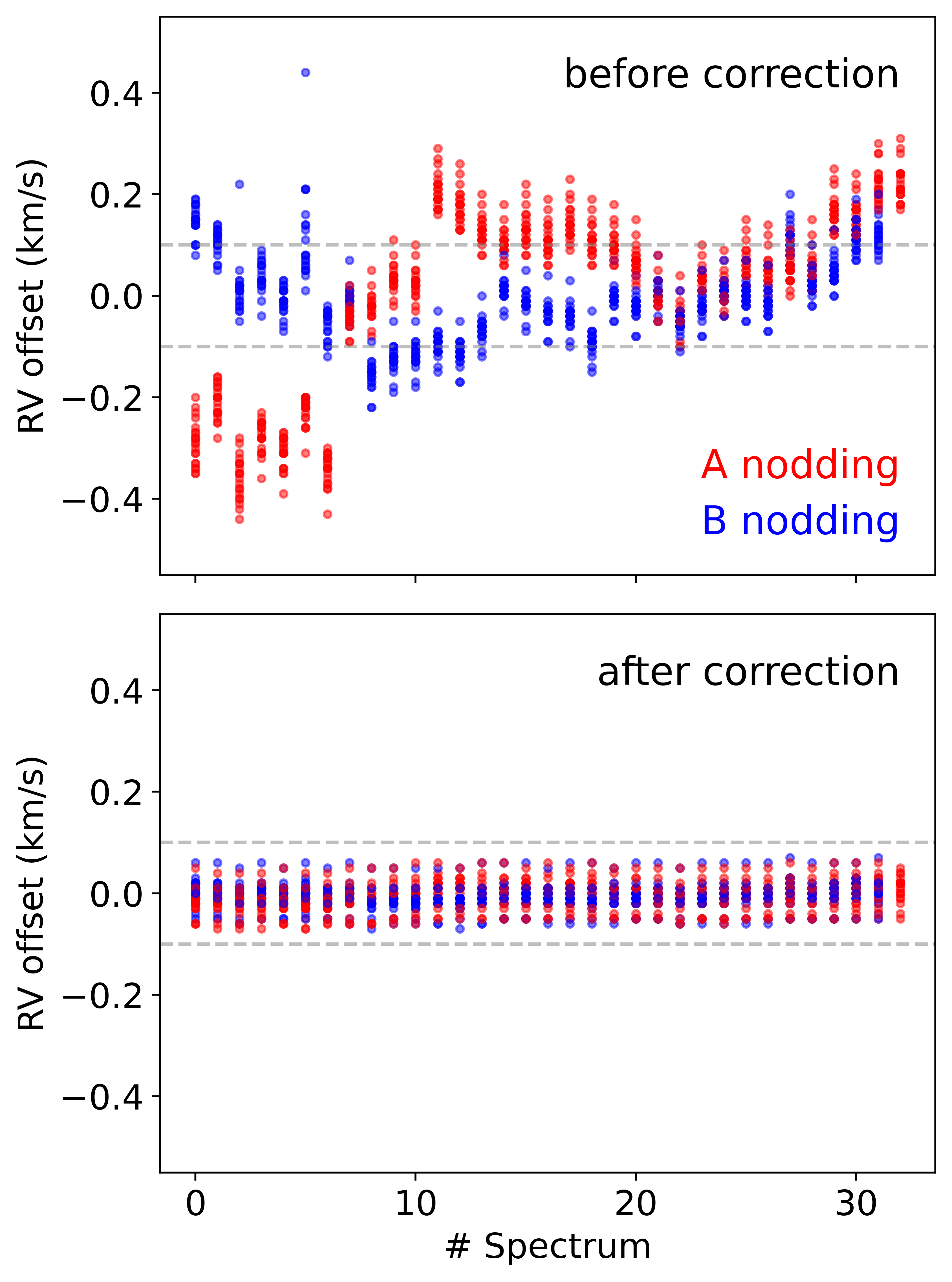}
        \caption{Radial velocity offsets before ({\it top panel}) and after ({\it bottom panel}) the alignment to the {\tt molecfit} wavelength solution. Each data point represents an individual spectrum and wavelength segment. Points corresponding to the A and B nodding positions are shown in red and blue, respectively. The $\pm0.1$\,km\,s$^{-1}$ interval around the average A and B wavelength solutions is indicated by the gray dashed lines.}
        \label{fig:wavelength-offset}
\end{figure}

\begin{figure}[h]
        \centering
        \includegraphics[width=\columnwidth]{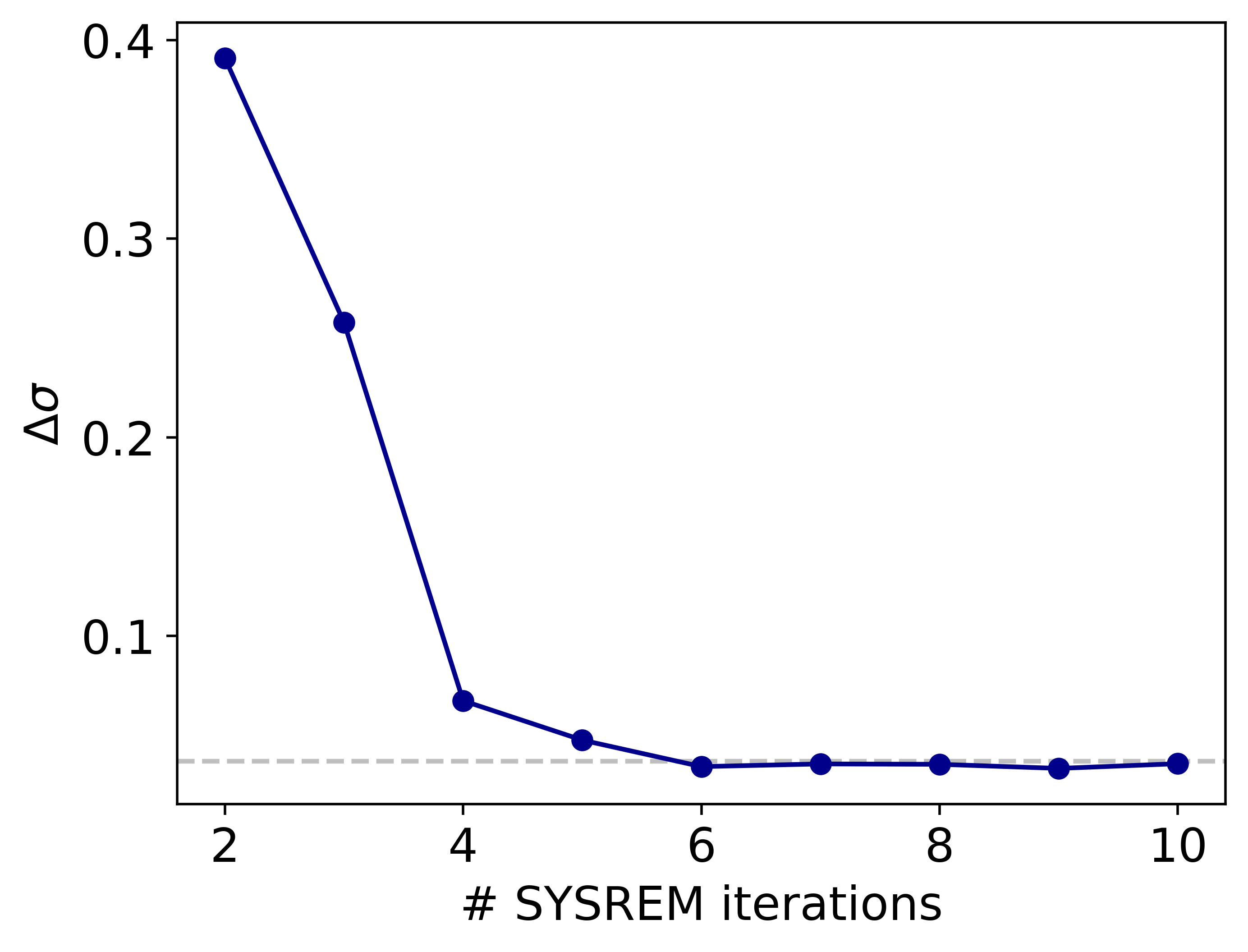}
        \caption{Relative change in the standard deviation of the residual spectral matrix between consecutive {\tt SYSREM} iterations. After six iterations, a plateau is reached, indicating that  {\tt SYSREM} no longer identifies significant correlated trends.}
        \label{fig:sysrem}
\end{figure}

\newpage

\begin{figure}[h]
        \centering
        \includegraphics[width=\columnwidth]{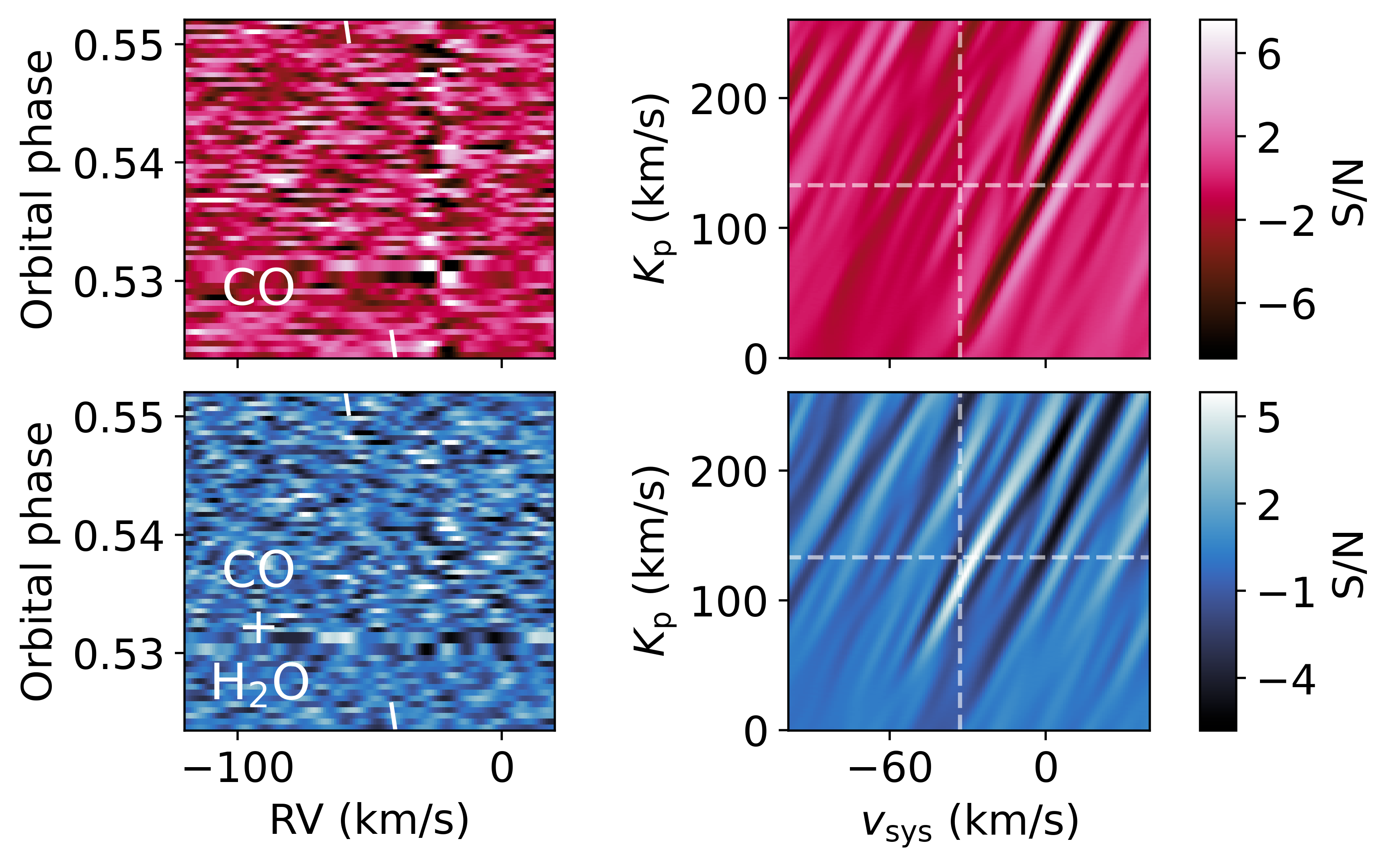}
        \caption{Same as in Fig.~\ref{fig:SN-CCF-model} but without masking the stellar CO residuals of the CCF maps when computing the S/N maps. For CO, the stellar residuals introduce strong artifacts into the S/N map. For the combined CO and H$_2$O signal, the artifacts are weaker, but still present, given the reduced S/N value compared to Fig.~\ref{fig:SN-CCF-model}.}
        \label{fig:SN-CCF-model-nomask}
\end{figure}
\onecolumn

\FloatBarrier

\newpage

\section{Posterior distributions}
\vspace{1cm}

\begin{figure*}[htbp]
        \includegraphics[width=\textwidth]{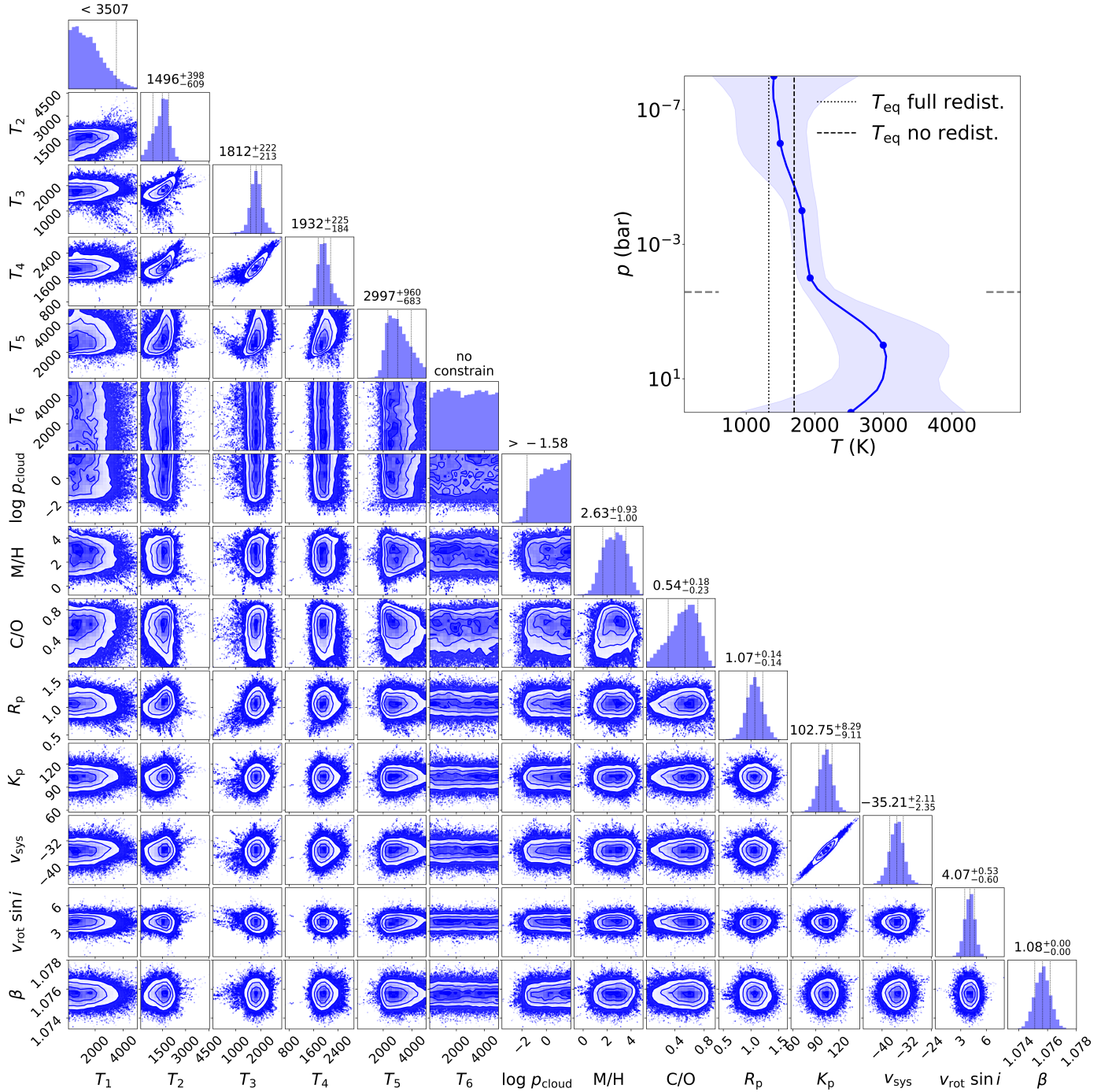}
        \caption{Retrieval results for \FiftyOnePegb. {\it Corner plot}: Posterior distributions and correlations between the retrieval parameters. The dashed vertical lines in each posterior distribution indicate the median and one-sigma intervals (16th, 50th, and 84th percentiles) for bounded parameters. For parameters with lower or upper limits, the two-sigma interval (5th or 95th percentile) is reported. {\it Top-right panel}: Median temperature curve with uncertainty intervals. The vertical dotted and dashed lines indicate the expected $T_\mathrm{eq}$ value of the planet, for the cases of full and no dayside to nightside heat redistribution, respectively. The horizontal gray lines indicate the retrieved lower pressure limit $p_\mathrm{cloud}$ of an optically thick cloud deck.}
        \label{fig:corner-plot}
\end{figure*}

\begin{figure*}[htbp]
        \centering
        \includegraphics[width=0.93\textwidth]{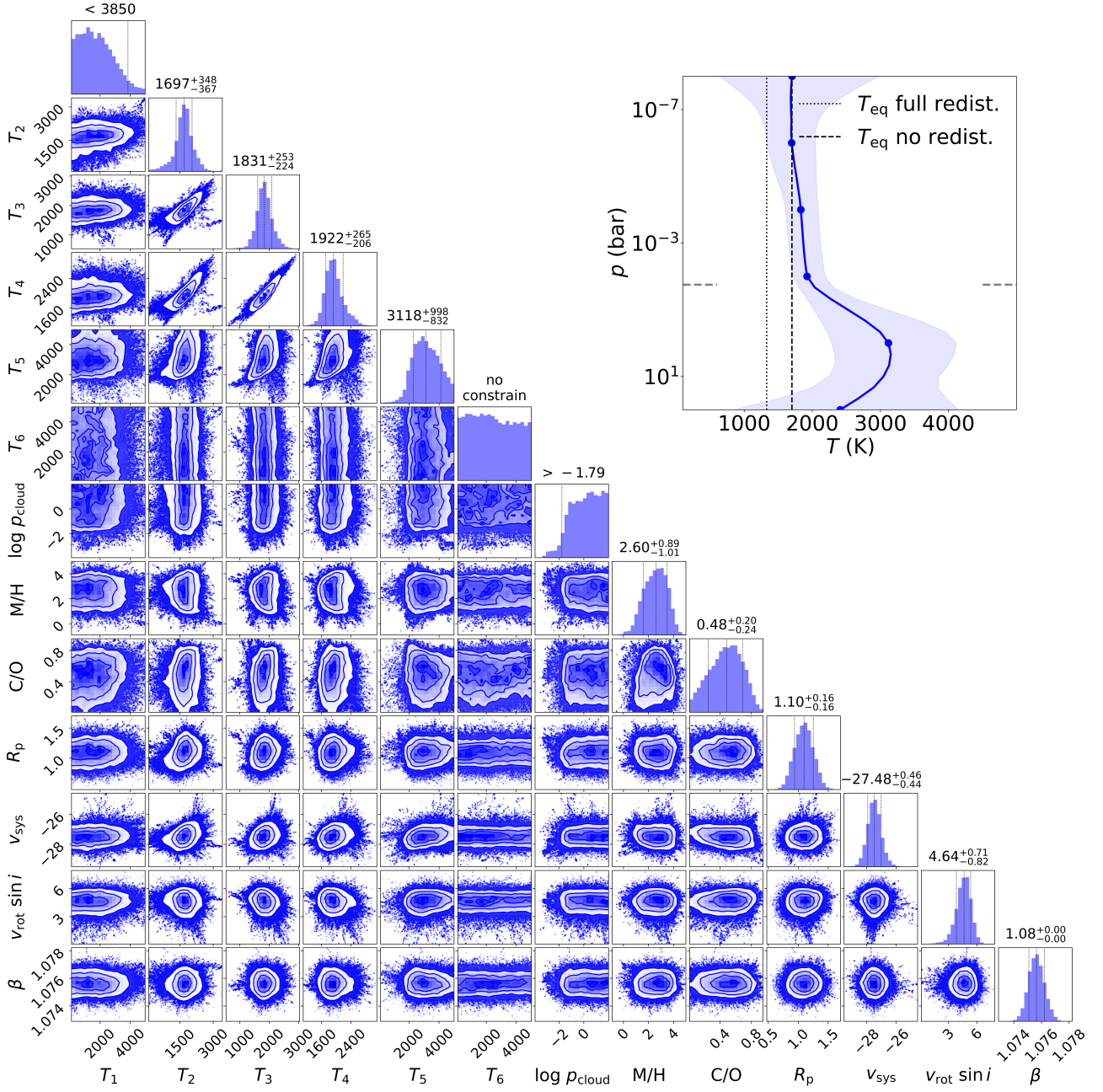}
        \caption{Same as Fig.~\ref{fig:corner-plot}, but with $K_\mathrm{p}$ fixed to 133\,km\,s$^{-1}$. All parameters except the systemic velocity are consistent with the free $K_\mathrm{p}$ retrieval. We note that that the spectral line broadening is slightly higher than in the free $K_\mathrm{p}$ retrieval. This indicates that a better alignment of the planetary spectral lines is achieved with the $K_\mathrm{p}$ value of approximately 103\,km\,s$^{-1}$ obtained by the free retrieval.}
        \label{fig:corner-plot-Kp133}
\end{figure*}

\twocolumn

\FloatBarrier

\begin{figure}[htbp]
        \centering
        \includegraphics[width=\columnwidth]{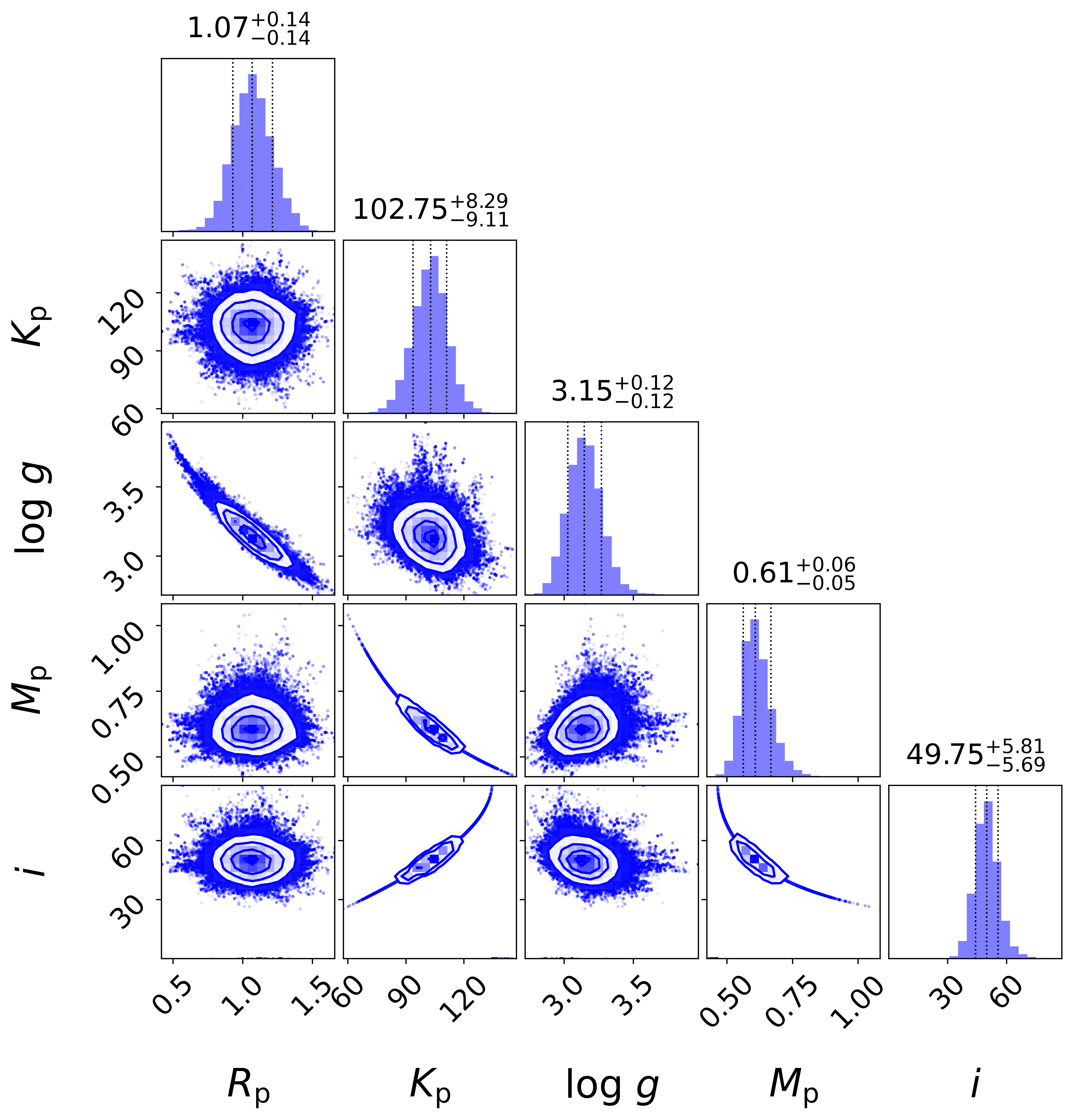}
        \caption{Same as Figs.~\ref{fig:corner-plot}, but for the derived planetary parameters. To illustrate the correlations between the free parameters $R_\mathrm{p}$ and $K_\mathrm{p}$, from which the derived parameters are obtained, we also include these free parameters in the figure.}
        \label{fig:corner-plot-derived-parameters}
\end{figure}

\end{document}